\documentclass[aps,prx,superscriptaddress,twocolumn,floatfix]{revtex4-1}

\usepackage{amsmath}
\usepackage{amssymb}
\usepackage{hhline}
\usepackage{float}
\usepackage{amssymb}
\usepackage{mathtools}
\usepackage{graphicx}
\usepackage[usenames,dvipsnames]{color}
\usepackage{bm}
\usepackage{hyperref}
\usepackage{url}
\usepackage[utf8]{inputenc}
\usepackage{subfigure}
\usepackage{slashed,bbm}
\usepackage{graphics,psfrag,epsfig}
\usepackage{dsfont}
\usepackage{setspace}
\usepackage{soul}
\usepackage{times}
\usepackage{amsfonts}
\usepackage{siunitx}

\usepackage{geometry}
\makeatletter
\newcommand*{\toccontents}{\@starttoc{toc}}
\makeatother

\usepackage{hyperref}
\usepackage{xcolor}
\usepackage{times}
\definecolor{dark-red}{rgb}{0.4,0.15,0.15}
\definecolor{dark-blue}{rgb}{0.15,0.15,0.4}
\definecolor{medium-blue}{rgb}{0,0,0.5}
\hypersetup{
colorlinks, linkcolor={dark-blue},
citecolor={dark-blue}, urlcolor={medium-blue}
}

\newcommand{\pdagger}{\phantom{\dagger}}
\newcommand{\diff}{\frac{d}{d\Lambda} }
\newcommand{\G}[4]{ \Gamma^{\Lambda}(#1, #2; #3, #4) }
\newcommand{\Gspatial}[6]{ \Gamma^{\Lambda}_{#1 #2}(#3, #4; #5, #6) }

\newcommand{\Gam}{\Gamma^{\Lambda}}

\newcommand{\Lim}[1]{\raisebox{0.5ex}{\scalebox{0.8}{$\displaystyle \lim_{#1}\;$}}}

\begin{document}

\title{Multiloop functional renormalization group approach to quantum spin systems}

\author{Dominik Kiese}
\email{dkiese@thp.uni-koeln.de}
\thanks{These two authors contributed equally}
\affiliation{Institute for Theoretical Physics, University of Cologne, 50937 Cologne, Germany}
\author{Tobias M\"uller}
\email{tobias.mueller@physik.uni-wuerzburg.de}
\thanks{These two authors contributed equally}
\affiliation{Institute for Theoretical Physics and Astrophysics, Julius-Maximilan's University of W\"urzburg, Am Hubland, 97074 W\"urzburg, Germany}
\author{Yasir Iqbal}
\affiliation{Department of Physics, Indian Institute of Technology Madras, Chennai 600036, India}
\author{Ronny Thomale}
\affiliation{Institute for Theoretical Physics and Astrophysics, Julius-Maximilan's University of W\"urzburg, Am Hubland, 97074 W\"urzburg, Germany}
\affiliation{Department of Physics, Indian Institute of Technology Madras, Chennai 600036, India}
\author{Simon Trebst}
\affiliation{Institute for Theoretical Physics, University of Cologne, 50937 Cologne, Germany}

\date{\today}

\begin{abstract}
	Renormalization group methods are well-established tools for the (numerical) investigation of the low-energy properties 
	of correlated quantum many-body systems, allowing to capture 
	their scale-dependent nature. 
	The functional renormalization group (FRG) allows to continuously evolve 
	a microscopic model action to an effective low-energy action as a function of decreasing energy scales via an exact functional flow equation, 
	which is then approximated by some truncation scheme to facilitate computation.
	Here, we report on our transcription of a recently developed multiloop truncation approach for electronic FRG calculations to the pseudo-fermion 
	functional renormalization group (pf-FRG) for interacting quantum spin systems.
	We discuss in detail the conceptual intricacies of the flow equations generated
	by the multiloop truncation, as well as essential refinements to the integration scheme for the resulting integro-differential 
	equations. 
	To benchmark our approach we analyze antiferromagnetic Heisenberg models
	on the pyrochlore, simple cubic and face-centered cubic lattice, 
	discussing the convergence of physical observables for higher-loop calculations
	and comparing with existing results where available.
	Combined, these methodological refinements systematically improve the pf-FRG approach to one of the numerical tools of choice
	when exploring frustrated quantum magnetism in higher spatial dimensions. 
\end{abstract}

\maketitle

\begingroup\renewcommand\thefootnote{\textsection}
\footnotetext{Equal contribution}
\endgroup

\section{Introduction}

The intriguing physics of quantum many-body systems often plays out on a multitude of scales. Archetypal examples include the spread of correlations on diverging length scales at phase transitions, the formation of coherent states of matter such as superconductivity at low temperatures, or the emergence of macroscopic entanglement in topological quantum liquids. \\
Capturing such diverse physics starting from simple microscopic models is a notoriously hard problem, since the most interesting phenomena manifest themselves solely at low temperatures and large system sizes. To establish a stringent connection between microscopic models and their effective low-energy, i.e., long-range physics, one often turns to 
renormalization group (RG) techniques that, by design, treat different scales iteratively rather than simultaneously, and thereby allow to evolve the original high-energy model description in an RG flow to an effective low-energy action \cite{POLCHINSKI1984269, Shankar1994}. \\
While the RG concept was originally developed in high-energy particle physics \cite{Stueckelberg1953}, its quick adaptation in the context of condensed matter physics and statistical physics has not only provided deeper understanding but also a multitude of applications and variations of the RG scheme. After Kadanoff's idea of a "block spin" renormalization group \cite{Kadanoff1966} to describe magnetic phase transitions, it was Wilson's numerical renormalization group (NRG)  \cite{NRGWilson, NRGReview} that led to the solution of the Kondo problem, i.e. the accurate, non-perturbative description of metallic conduction electrons coupled to a magnetic impurity below the Kondo temperature $T_{k}$ and the explanation of the finite electrical resistivity that these systems exhibit at ultralow temperatures \cite{Kondo}. The density matrix renormalization group (DMRG) developed by White \cite{DMRGWhite} to capture the formation of entanglement in the ground states of quantum many-body systems, has basically solved the one-dimensional interacting quantum many-body problem \cite{DMRGReview}. Its application to two-dimensional systems \cite{DMRG2D} and its generalization to tensor network approaches \cite{TensorNetworks} is one of the most active developments in contemporary computational physics. \\
When it comes to systems of interacting electrons in two and three spatial dimensions, a particularly appealing flavor of the renormalization group is the functional renormalization group (FRG) \cite{POLCHINSKI1984269, Wetterich1993}. 
This approach, which will be the foundation of this manuscript, is based on an infinite hierarchy of ordinary integro-differential equations; they govern the evolution of $n$-particle Green's functions or vertices controlled by a flow parameter $\Lambda$ (usually chosen as an infrared cutoff). This allows to systematically derive effective low-energy actions for interacting electron problems, and is routinely employed to elucidate the pairing mechanism in novel superconductors, or other kinds of Fermi surface instabilities \cite{MetznerReview, Thomalereview}. In practice, unless for the exactly solvable model originally studied by Polchinski~\cite{POLCHINSKI1984269}, the FRG necessitates approximations imposed on the coupled integro-differential flow equations in order to render their numerical solution feasible. \\
First, one needs to truncate their hierarchy to a level which covers the physics of interest, but is still amenable to semi-analytical or numerical approaches. Most often one considers $n$-point functions with $n \leq 4$ and treats higher-order contributions only to a small extent. Truncations which completely neglect these Green's functions are especially justified when the bare interactions are weak, and corrections to the flow thus are presumably small. As it turns out, FRG studies of itinerant fermion models have reached a remarkable degree of precision for determining ground state phase diagrams of, e.g., the Hubbard model at, and even away, from half-filling \cite{MetznerReview, Hubbard_1, Hubbard_2}. \\
Second, there exists no unique way of implementing the RG parameter $\Lambda$ into the generator of the vertices. Since the flow equations are only used in their truncated form, it naturally introduces a dependence of the results on the choice of regulator function. For FRG, this has often led to a certain inherent dependence of quantitative predictions on the actual choice of regularization.  \\
Recently, the {\sl multiloop} truncation \cite{Kugler_1, Kugler_2} of the FRG flow equations (ml-FRG) has been developed to overcome some of these shortcomings. This is done by iteratively advancing the flow of the two-particle vertex to arbitrary orders in the bare interaction, until convergence to the first-order parquet equation \cite{Diatlov-1957, *Bychkov-1966, *Roulet-1969, Abrikosov-1965} is reached. Thereby, one recovers an independence of the choice of regulator, while simultaneously keeping the additional numerical cost at a manageable level. For itinerant electron systems, this approximation has been found to improve the outcome of the FRG calculations, e.g. allowing for quantitative agreement with determinant quantum Monte Carlo simulations of the two dimensional Hubbard model \cite{HilleQuantitative}. For intermediate interaction strengths, a high degree of convergence in the number of loops $\ell$ was found to be reached already at $\ell \approx 8$ \cite{Tagliavini}, with the numerical effort scaling linearly in $\ell$. \\
In this manuscript, we transcribe the multiloop scheme to the pseudo-fermion FRG (pf-FRG) approach to quantum spin systems \cite{ReutherOrig, Reuther-2011a, LargeS, LargeN}. Based on a decomposition of spin operators into fermionic partons \cite{Abrikosov-1965}, this adaptation of the FRG scheme allows to study the physics of frustrated quantum magnets in two \cite{ReutherOrig, Reuther-2011a, Reuther-2011b, Reuther-2011c, ReutherKitaev, Reuther-2012, Reuther-2014a, Suttner-2014, Reuther-2014b, Rousochatzakis-2015, Iqbal-2015, Iqbal-2016b, Iqbal-2016a, Buessen-2016, Laubach-2017, Keles-2018a, Keles-2018b, Iida-2020, KieseSpinValley} and three spatial dimensions \cite{Balz-2016, Iqbal3D, Buessen-2016, Iqbal-2017, BuessenDiamond, Iqbal-2018a, MuellerPyrochlore, MuellerBCC, KieseFCC, Niggemann-2019, Ghosh-2019, Chillal-2020}, which are commonly beyond the reach of other numerical quantum many-body schemes. \\ 
On a technical level, our multiloop pf-FRG approach introduced here is a transcription of the multiloop weak coupling implementations mentioned above. Besides certain subtleties that result from the bilocal parametrization of the two-particle vertex in real space, our technical formulation of the multiloop equations is in agreement with earlier studies \cite{Kugler_1}. Furthermore, we have implemented a characterization of the high-frequency structure of vertex functions which fully captures their asymptotic behavior \cite{WentzellAsymptotics, Tagliavini, HilleQuantitative, HillePseudogap}, in order to attenuate numerical artifacts at higher loop orders, and to stabilize the flow of all dressed couplings. \\
We benchmark our method by applying the ml-FRG to Heisenberg models on various three dimensional lattices, subject to different levels of frustration. For the antiferromagnets on the pyrochlore and cubic lattice we distill the impact of higher loops on the signatures of the respective ground states, i.e. the symmetry-preserving Coloumb spin liquid phase for the former \cite{MuellerPyrochlore} and the symmetry-broken N\'eel state for the latter \cite{SandvikCubic, Iqbal3D}. We then add a finite third-nearest neighbor coupling $J_3$ to the antiferromagnetic nearest-neighbor Heisenberg model on the simple cubic lattice and map out the phase diagram both in the unfrustrated regime $J_3 / J_1 > 0$, as well as for mildly frustrated $J_3 / J_1 < 0$. As a last step of exemplary numerical analysis, we study the rich phase diagram of the $J_1-J_2$ Heisenberg model on the fcc lattice, featuring spin liquid candidates with sub-extensively degenerate ground state manifolds as well as magnetically ordered phases \cite{KieseFCC, finnish_93, Haar-1962, Alexander-1980, IqbalFCC}. \\
The manuscript is structured as follows. In Section II, we review the conventional formulation of pf-FRG as put forward in Refs. \cite{ReutherOrig, Reuther-2011a, Iqbal3D, LargeS, LargeN, BuessenOffDiag}. We further proceed by highlighting the parametrization of the high frequency structure of the two-particle vertex \cite{WentzellAsymptotics} and the multiloop truncation. In Section III, we discuss our refinements of the numerical implementation of the pf-FRG procedure. Finally, for Section IV, we present our benchmark results for Heisenberg models on the pyrochlore, cubic and face-centered cubic lattice. In Section V, we conclude that the multiloop pf-FRG promises to rise up as one of the few numerical approaches available today that are capable of analyzing quantum magnetism in higher dimensions. We further speculate on the next potential methodological extensions and improvements of pf-FRG which can use our work as a reference point in terms of conceptual implementation and numerical performance. 

\section{Method}

In this section, we briefly review the conventional formulation of pf-FRG as put forward in earlier studies \cite{ReutherOrig, ReutherKitaev, BuessenOffDiag, BuessenThesis, BuessenDiamond, LargeS, LargeN, MuellerBCC, MuellerPyrochlore, KieseFCC, KieseSpinValley, Iqbal3D}, before we continue with a discussion of the methodological extensions which are subject to this manuscript.

\subsection{Conventional pf-FRG}

\begin{figure*}[t]
	\centering
	\includegraphics[width = 1.0\linewidth]{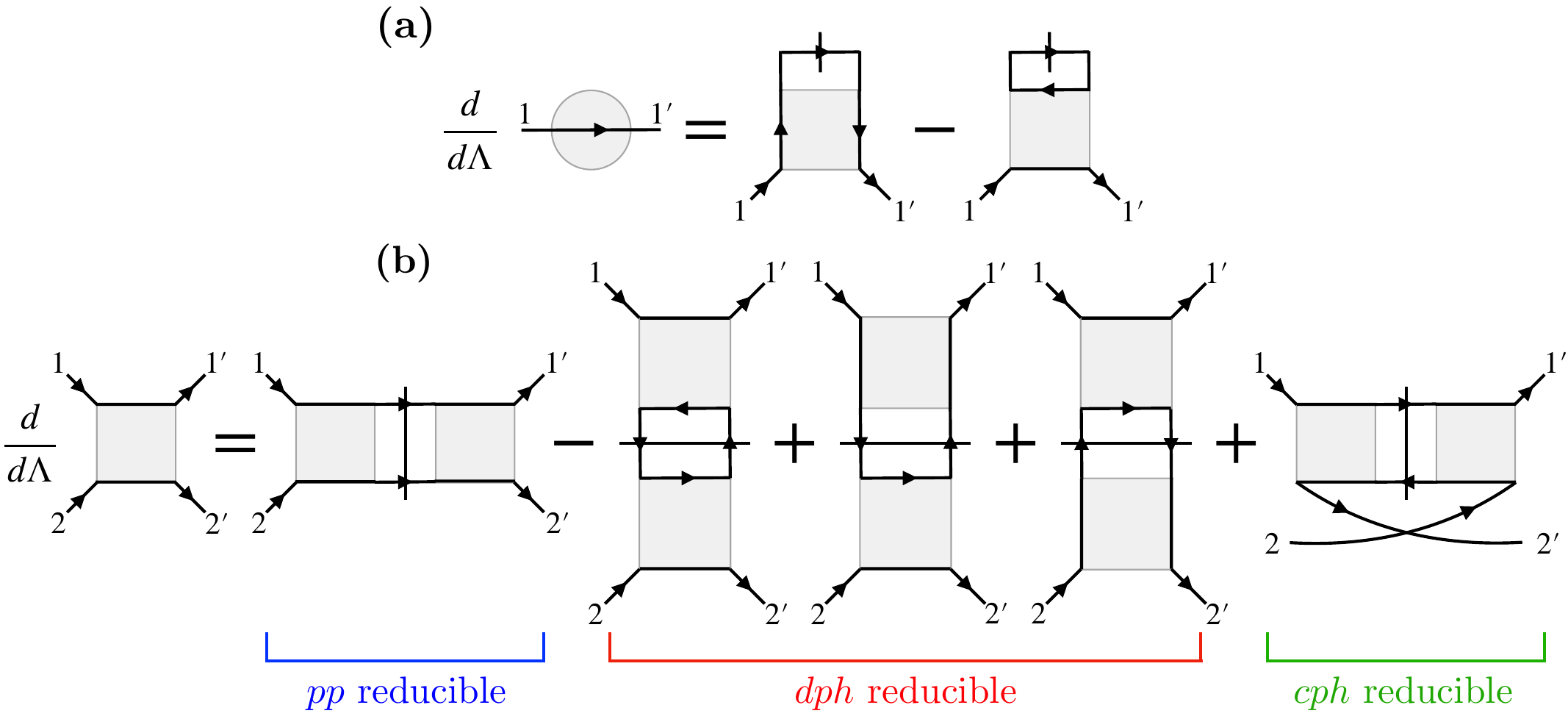}
	\caption{\textbf{Diagrammatic representation of the flow equations Eq.~\eqref{self} and Eq.~\eqref{2Pgeneral} in terms of bilocal vertices} where conserved lattice indices are indicated by thick black lines. The self-energy flow (a) is decomposed into a local Fock diagram and a non-local Hartree term which contains a summation over the full lattice. The two-particle vertex flow (b) can be written as a sum of five terms differing either in their two-particle reducibility or, in the case of the three diagrams reducible in the $dph$ channel, in their spatial structure. Slashed lines denote pairs of differentiated propagators.}
	\label{fig:flowequation}
\end{figure*}

Our starting point is a spin-$1/2$ Heisenberg model of $SU(2)$ spins
\begin{align}
	\mathcal{H} = \sum_{ij} J_{ij} \mathbf{S}_i \mathbf{S}_j \,,
	\label{ham}
\end{align}
on a lattice with sites $i, j$ subject to real exchange couplings $J_{ij}$. The spin operators are represented in terms of complex pseudo-fermions $f^{(\dagger)}_{i \alpha}$ with $\alpha \in \{\uparrow, \downarrow \}$~\cite{Abrikosov-1965}, i.e.,
\begin{align}
S^{\mu}_i = \frac{1}{2} \sum_{\alpha, \beta} f^{\dagger}_{i \alpha} \sigma^{\mu}_{\alpha \beta} f^{\pdagger}_{i \beta}\,,
\label{partons}
\end{align}
where $\sigma^{\mu}_{\alpha \beta}$ for $\mu \in \{x, y, z\}$ denote Pauli matrices. While this results in a purely quartic Hamiltonian which can directly be treated by established functional RG techniques \cite{MetznerReview}, the pseudo-fermion representation of the spin algebra is a priori not isomorphic to the original spin-$1/2$ representation, since the dimensions of the Hilbert spaces of pseudo-fermions ($d = 4$) and spin operators ($d = 2$) differ. However, unphysical Fock states with net zero spin can be projected out by an additional particle number constraint $\sum_{\alpha} f^{\dagger}_{i\alpha} f^{\pdagger}_{i \alpha} = 1$, which has to be fulfilled on all lattice sites individually. In practice, this constraint is only enforced on average by an explicit implementation of particle-hole symmetry on the level of irreducible vertex functions \cite{LargeS, BuessenOffDiag, KieseSpinValley} and can directly be tested by computation of the average $\langle f^{\dagger}_{i\alpha} f^{\pdagger}_{i \alpha} \rangle$ (cf. Appendix~\ref{appendix:halffilling}). Furthermore, the impact of occupation number fluctuations can be checked for by implementing local level repulsion terms which gap out the unphysical states \cite{LargeS, KieseSpinValley, BuessenDiamond, LMU-Group}. \\
Due to the absence of kinetic contributions, the free propagator for the pseudo-fermion Hamiltonian takes the simple form
\begin{align}
	G_0(w) = (iw)^{-1} \,,
\end{align}
diagonal in real and spin space, where $w$ is a fermionic Matsubara frequency. Similar to other flavors of FRG, a regulator function $\Theta^{\Lambda}(w)$ is introduced to cut off infrared divergencies in a controlled manner: for $\Lambda \to \infty$ the product of full propagator and regulator vanishes, while the original system is recovered for $\Lambda \to 0$. Here we choose
\begin{align}
    \Theta^{\Lambda}(w) = 1 - e^{-w^2 / \Lambda^2} \,.
    \label{regulator}
\end{align}
The functional renormalization group equations for the $n$-particle vertices then correspond to an interpolation between the simple limit where vertices collapse to the bare interaction $J_{ij}$ and the physical limit of vanishing cutoff. Although these equations are in principle exact, the full hierarchy of integro-differential equations is not closed, rendering approximations necessary in attempts to seek for its solution. \\
Previous implementations of pf-FRG \cite{ReutherOrig, LargeS, LargeN, Iqbal3D, BuessenThesis, KieseSpinValley} have made extensive use of the Katanin scheme which truncates the FRG equations after the two-particle vertex while simultaneously approximating contributions from the three-particle vertex by a self-energy feedback in the two-particle vertex flow. After truncation, the flows for the self-energy $\Sigma^{\Lambda}$ and the two-particle vertex $\Gamma^{\Lambda}$ read
\begin{align}
	\diff \Sigma^{\Lambda}(1) &= -\frac{1}{2\pi} \sum_2 \G{1}{2}{1}{2} S^{\Lambda}(2) \notag \\
	&\equiv - \big{[} \Gam \circ S^{\Lambda} \big{]}_{\Sigma}
	\label{self}
\end{align}
\begin{align}
	\diff &\G{1'}{2'}{1}{2} = \notag\\
	&\frac{1}{2\pi} \sum_{3, 4} \big{[} \G{3}{4}{1}{2} \G{1'}{2'}{3}{4} \notag  \\ 
	&- \G{1'}{4}{1}{3} \G{3}{2'}{4}{2} - (3 \leftrightarrow 4)                             \notag  \\ 
	&+ \G{2'}{4}{1}{3} \G{3}{1'}{4}{2} + (3 \leftrightarrow 4) \big{]}                           \notag  \\ 
	&\times G^{\Lambda}(3) S^{\Lambda}(4) \,,
	\label{2Pgeneral}
\end{align}
where the compound indices comprise a lattice and a spin index as well as a Matsubara frequency e.g. $1 = (i_1, \alpha_1, w_1)$. Conjugate Grassmann fields are discriminated by primes attached to the respective index, where $1'$ indicates an outgoing and $1$ an incoming fermionic parton. Furthermore, $S^{\Lambda} \equiv -\diff G^{\Lambda}|_{\Sigma^{\Lambda} = \text{const.}}$ is the single-scale propagator.  Note that due to local $U(1)$ and global $SU(2)$ symmetry of the Hamiltonian in Eq.~\eqref{ham} the self-energy as well as the dressed propagators are diagonal in real and spin space. The Katanin truncation now amounts to the replacement
\begin{align}
	S^{\Lambda} \to -\diff G^{\Lambda}
	\label{SubKatanin}
\end{align}
in the $\Gam$ flow. In this form, the pf-FRG equations become equivalent to mean-field gap equations in the limit of large spin length $S$, where they collapse to a mere resummation of RPA diagrams, as well as large dimension of the spin algebra $N$ \cite{LargeS, BuessenThesis}, where only crossed particle-hole diagrams remain. \\
Transitions into phases with broken symmetries become visible in pf-FRG by an instability (indicated by a kink, cusp or divergence) in the flowing spin-spin correlation
\begin{align}
	\chi^{\Lambda}_{ij}(iw = 0) = \int_{0}^{\beta} d\tau \langle T_{\tau} S^{\mu}_{i}(\tau) S^{\mu}_{j}(0) \rangle^{\Lambda} \,,
\end{align}
where the renormalization has to be stopped to still extract sensible results. Here, $T_{\tau}$ is the imaginary time ordering operator and $\mu \in \{x, y, z\}$ can be chosen arbitrarily due to spin rotation invariance of Eq.~\eqref{ham}. For long-range ordered states, the momentum $\mathbf{k}$ for which the susceptibility (i.e. the Fourier transform of $\chi_{ij}$) is most dominant, characterizes the respective type of order. The absence of a flow breakdown is, on the other hand, associated with putative spin liquid phases.

\subsection{Asymptotic frequency parametrization}

\begin{figure*}
	\centering
	\includegraphics[width = 1.0\linewidth]{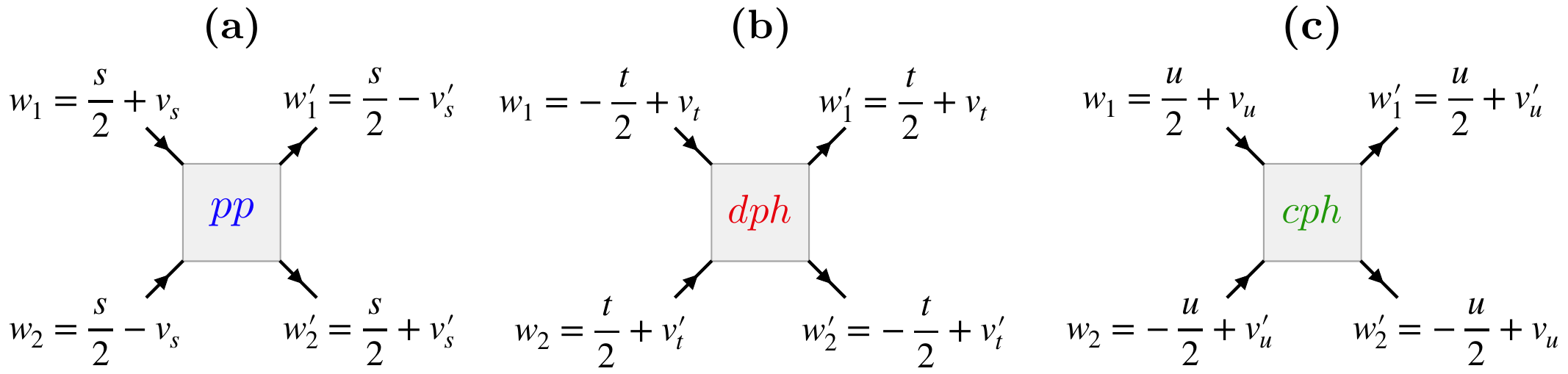}
	\caption{\textbf{Symmetrized frequency parametrization of the two-particle vertex channels.} Shifting all arguments by half a transfer frequency allows for a more convenient implementation of symmetries on the level of vertex functions  (see Appendix~\ref{appendix:symmetries} for more details).}
	\label{fig:symmetric_parametrization}
\end{figure*}

For the $T = 0$ implementation of pf-FRG, the spectrum of Matsubara frequencies becomes continuous and vertices need to be discretized on a finite number of frequency mesh points in order to compute a numerical RG flow. Moreover, a crucial ingredient for the solution of the truncated set of equations is the integration of products of Green's functions in frequency space during the evaluation of the inner sums in Eq.~\eqref{2Pgeneral}. Hence, numerical computations with limited resolution need to capture all relevant features of the vertices to obtain robust results. We employ an established parametrization scheme, which sorts all diagrams that may become finite during the flow into one of four classes and thereby tracks the high-frequency structure of the two-particle vertex in an efficient manner \cite{WentzellAsymptotics}. We start by grouping the contributions in Eq.~\eqref{2Pgeneral} into three channels, which differ in their two-particle reducibility, i.e the way in which external legs are assigned to vertices after cutting the two propagators in the respective diagrams. In this sense, the first term is particle-particle ($pp$) reducible, the second one direct particle-hole ($dph$) reducible and the last one crossed particle-hole ($cph$) reducible. Equation \eqref{2Pgeneral} with the Katanin substitution \eqref{SubKatanin} can therefore be compactly stated as
\begin{align}
	\diff \Gam &= \dot{g}_{pp}^{\Lambda} + \dot{g}_{dph}^{\Lambda} + \dot{g}_{cph}^{\Lambda} \\ 
	\dot{g}_{c}^{\Lambda} &\equiv \big{[} \Gam \circ \partial_{\Lambda} (G^{\Lambda} \times G^{\Lambda}) \circ \Gam \big{]}_{c} \,,
	\label{channeldecomp}
\end{align}
where the precise definitions of the channels are given in Appendix \ref{appendix:channel_definitions}. As a consequence of imaginary time translation invariance and therefore Matsubara frequency conservation, each term can be associated with a specific bosonic frequency, corresponding to the energy transferred through the internal loop: $pp$ with $s = w_{1'} + w_{2'}$, $dph$ with $t = w_{1'} - w_{1}$ and $cph$ with $u = w_{1'} - w_{2}$. If the frequency dependence of the channels is projected onto the respective transfer frequency, two independent fermionic frequency arguments remain to be determined, with our choice displayed in [Fig.~\ref{fig:symmetric_parametrization}]. This specific parametrization simplifies the internal symmetries of the channels under frequency inversions and exchange of fermionic frequencies (cf. Appendix~\ref{appendix:symmetries}). \\
The diagrams contributing to each channel are classified according to the number of external arguments, which enter the internal summations \cite{WentzellAsymptotics}, i.e.
\begin{align}
	\dot{g}_{c}^{\Lambda}(w_c, v_c, v'_c) &= K_{1c}^{\Lambda}(w_c) + K_{2c}^{\Lambda}(w_c, v_c)  \notag \\
	&+ \bar{K}_{2c}^{\Lambda}(w_c, v'_c) + K_{3c}^{\Lambda}(w_c, v_c, v'_c) \,,
	\label{kernels}
\end{align}
for $c \in \{pp, dph, cph\}$. Note that we have only stated the frequency dependence explicitly while suppressing site and spin indices. Each kernel captures a certain asymptotic limit of the channels, since they decay to zero if one of their respective arguments is taken to infinity \cite{WentzellAsymptotics}. This can be seen by recalling that the full propagator effectively scales as $1/(iw)$ for large Matsubara frequencies. In this regard, one gains, in principle, explicit access to the asymptotic behavior of all contributions allowing us to model different diagrams more effectively in numerical calculations, where only a finite number of frequencies can be used. For computational purposes, however, it is far more advantageous to define new kernels
\begin{align}
    Q_{1c}^{\Lambda}(w_c) &\equiv \Lim{|v_c|, |v'_c| \to \infty} \dot{g}_{c}^{\Lambda}(w_c, v_c, v'_c) \notag \\
    Q_{2c}^{\Lambda}(w_c, v_c) &\equiv \Lim{|v'_c| \to \infty} \dot{g}_{c}^{\Lambda}(w_c, v_c, v'_c) \notag \\
    \bar{Q}_{2c}^{\Lambda}(w_c, v'_c) &\equiv \Lim{|v_c| \to \infty} \dot{g}_{c}^{\Lambda}(w_c, v_c, v'_c) \notag \\
    Q_{3c}^{\Lambda}(w_c, v_c, v'_c) &\equiv \dot{g}_{c}^{\Lambda}(w_c, v_c, v'_c)\,,
\end{align}
where the limits are either performed numerically, by setting the respective frequency to a large value, or by scanning the boundaries of $Q_{3c}$ after evaluating the right-hand side of the flow equations. In the latter case, the asymptotic classes, though they can individually be extracted in each stage of the flow, only serve as efficient numerical buffers for constant extrapolations beyond the domain where $Q_{3c}$ has been discretized. The new functions are related to the old kernels by
\begin{align}
	Q_{1c}^{\Lambda}(w_c) &= K_{1c}^{\Lambda}(w_c) \notag \\
	Q_{2c}^{\Lambda}(w_c, v_c) &= K_{1c}^{\Lambda}(w_c) + K_{2c}^{\Lambda}(w_c, v_c) \notag \\
	\bar{Q}_{2c}^{\Lambda}(w_c, v'_c) &= K_{1c}^{\Lambda}(w_c) + \bar{K}_{2c}^{\Lambda}(w_c, v'_c) \notag \\
	Q_{3c}^{\Lambda}(w_c, v_c, v'_c) &= K_{1c}^{\Lambda}(w_c) + K_{2c}^{\Lambda}(w_c, v_c) \notag \\ &+ \bar{K}_{2c}^{\Lambda}(w_c, v'_c) + K_{3c}^{\Lambda}(w_c, v_c, v'_c) \,.
\end{align}
Keeping only these sums, one significantly reduces the number of memory accesses in a numerical implementation of the method, since for a given set of frequency arguments, only one function $Q^{\Lambda}$ needs to be accessed instead of multiple kernels $K^{\Lambda}$. Another advantage of this definition is that the additional cost of extracting the asymptotic functions after computing limits of the flow \cite{WentzellAsymptotics} is avoided.

\subsection{Multiloop extension}

In the context of FRG for itinerant fermions \cite{Tagliavini, HilleQuantitative, HillePseudogap}, it has been shown that an extended truncation, dubbed 'multiloop' scheme, leads to a substantial improvement of functional RG calculations by 1) restoring independence of the choice of regulator function for $\Lambda \to 0$ \cite{Kugler_1, Kugler_2} and 2) generation of all two-particle reducible (parquet) diagrams, which can be computed at a manageable numerical cost. This multiloop FRG (ml-FRG) scheme is based on the parquet equations i.e the Schwinger-Dyson equation (SDE) connecting the self-energy to the two-particle vertex and the Bethe-Salpeter equations (BSEs) for the two-particle reducible channels, which compactly written read
\begin{align}
    \Sigma &= \big{[} \big{(} \Gamma_0 + \big{[} \Gamma_0 \circ (G \times G) \circ \Gamma \big{]}_{pp} \big{)} \circ G  \big{]}_{\Sigma} \\ 
    g_c &= \big{[} \big{(} \Gamma - g_c \big{)} \circ (G \times G) \circ \Gamma \big{]}_{c} \,.
\end{align}
Note that we have already applied the well known parquet approximation (PA), substituting the fully irreducible vertex with the bare vertex $\Gamma_{0}$. To construct from the parquet equations (in the PA) the ml-FRG flow, one regularizes the propagators as in Eq.~\eqref{regulator}. In consequence the SDE and BSEs become scale dependent and can be put into differential form by taking derivatives with respect to $\Lambda$ on both sides of the equation. The multiloop flow in a channel $g_c$ can then be computed via an iterative scheme which reads \cite{Kugler_1, Kugler_2}
\begin{widetext}
    \begin{align}
        \dot{g}_c & = \sum_{\ell \geq 1} \dot{g}^{(\ell)}_{c} \\
        \dot{g}^{(1)}_{c} &= \big{[} \Gamma \circ \partial_{\Lambda} (G \times G) \circ \Gamma \big{]}_{c} \\ 
        \dot{g}^{(2)}_{c} &= \big{[} \dot{g}^{(1)}_{\bar{c}} \circ (G \times G) \circ \Gamma \big{]}_{c} + \big{[} \Gamma \circ (G \times G) \circ \dot{g}^{(1)}_{\bar{c}} \big{]}_{c} \equiv \dot{g}^{(2), L}_{c} + \dot{g}^{(2), R}_{c} \\
        \dot{g}^{(\ell \geq 3)}_{c} &= \dot{g}^{(\ell), L}_{c} + \big{[} \dot{g}^{(\ell - 1), R}_{c} \circ (G \times G) \circ \Gamma \big{]}_{c} + \dot{g}^{(\ell), R}_{c} \\ 
        &= \dot{g}^{(\ell), L}_{c} + \big{[} \Gamma \circ (G \times G) \circ \dot{g}^{(\ell - 1), L}_{c} \big{]}_{c} + \dot{g}^{(\ell), R}_{c} \\ 
        &\equiv \dot{g}^{(\ell), L}_{c} + \dot{g}^{(\ell), C}_{c} + \dot{g}^{(\ell), R}_{c} \,,
        \label{mFRG_scheme}
    \end{align}
\end{widetext}
where we have defined the left, right and central part of the $\ell$ loop contribution. The flow equation for the self-energy Eq.~\eqref{self} is in principle exact, at least given an exact two-particle vertex $\Gam$. One computes, however, an approximate RG flow for the latter, such that additional corrections become necessary \cite{Kugler_2, HillePseudogap, HilleQuantitative}. The ml-FRG flow for the self-energy then reads 
\begin{align}
    \dot{\Sigma} &= \dot{\Sigma}_{0} + \dot{\Sigma}_{1} + \dot{\Sigma}_{2} \\ 
    \dot{\Sigma}_{0} &= -\big{[} \Gamma \circ S \big{]}_{\Sigma} \\
    \dot{\Sigma}_{1} &= \big{[} \sum_{\ell \geq 3} \big{(} \dot{g}^{(\ell), C}_{pp} + \dot{g}^{(\ell), C}_{cph}\big{)} \circ G \big{]}_{\Sigma} \\
    \dot{\Sigma}_{2} &= \big{[} \Gamma \circ (G \times \dot{\Sigma}_{1} \times G) \big{]}_{\Sigma} \,.
\end{align}
Since the flow of the vertex requires the self-energy derivative, which itself builds on the central parts of the particle-particle and crossed particle-hole channels, one usually computes the vertex corrections using only the standard expression $\dot{\Sigma} = \dot{\Sigma}_{\text{0}}$ and accounts for self-energy corrections $\dot{\Sigma}_{\text{1}}, \dot{\Sigma}_{\text{2}}$ afterwards. The revised value for $\dot{\Sigma}$ can in turn be used to recompute the vertex corrections until convergence is reached. In this work, however, these numerically expensive self-energy loops are not considered, as the self-energy corrections already turn out to be small during the flow.

\section{Numerical implementation}

In order to treat the closed set of integro-differential equations forming the truncated pf-FRG equations, we have to introduce a few more approximations to both the infinite real space lattice, and the continuous Matsubara frequencies, to  make them numerically tractable.

\subsection{Finite lattice graphs}

The parton decomposed spin operators \eqref{partons} are invariant under local $U(1)$ transformations $f^{(\dagger)}_{i\alpha} \to e^{\pm i\phi} f^{(\dagger)}_{i\alpha}$ implying conservation of the number of spinons per lattice site. The site dependence of the two-particle vertex can therefore be efficiently reduced by the bilocal parametrization \cite{BuessenOffDiag}
\begin{align}
    \G{1'}{2'}{1}{2} &= \Gamma^{\Lambda =}_{i_1 i_2}(1', 2'; 1, 2)      \delta_{i_{1'} i_1} \delta_{i_{2'} i_2} \notag \\
                     &+ \Gamma^{\Lambda \times}_{i_1 i_2}(1', 2'; 1, 2) \delta_{i_{1'} i_2} \delta_{i_{2'} i_1} \,,
    \label{bilocal}
\end{align}
where vertices with crossed fermion lines $\Gamma^{\Lambda \times}_{i_1 i_2}$ can be replaced by vertices with parallel fermion lines $\Gamma^{\Lambda =}_{i_1 i_2}$ (or vice versa) by making use of the crossing symmetry $\G{1'}{2'}{1}{2} = -\G{2'}{1'}{1}{2}$. We therefore focus only on vertices with parallel lines in the following and drop the additional superscript "$=$" for brevity [Fig.~\ref{fig:flowequation}]. \\
In addition, by treating all sites as symmetry equivalent, the site dependence of the self-energy can be entirely discarded, while lattice symmetries can be employed to obtain an effective dependence on a single site $i^{*}_1$ for the two-particle vertex i.e. $\Gamma^{\Lambda}_{i_1 i_2} \rightarrow \Gamma^{\Lambda}_{i^{*}_1 i_0}$. Here $i_0$ is a fixed reference site, taken to be invariant under point group symmetries, and $i^{*}_1$ is the image of $i_1$ for $i_2$ mapped to $i_0$. Given a unitcell of the lattice, our code automatically performs this symmetry reduction, by explicitly computing transformations, which leave the lattice invariant. Finally, vertices are truncated if the bond distance $d(i^{*}_1, i_0)$ exceeds a threshold $L$, which amounts to artificially introducing a maximal correlation length. In this manuscript, we choose $L = 6$ to keep the numerical effort for the multiloop truncation in conjunction with the three-dimensional lattices of interest at a manageable level.

\subsection{Matsubara frequency discretization and integration}

The pf-FRG flow equations have been derived in the $T = 0$ limit, where Matsubara frequencies become continuous and internal summations are promoted to integrals. In order to solve the flow equations numerically, one therefore has to make an appropriate choice both for the integration algorithm as well as the discretization of the vertices on a finite grid. To this end, one should carefully consider the interplay between the choice of regulator function, the propagators and the vertices.
\begin{figure}
	\centering
	\includegraphics[width = 1.0\columnwidth]{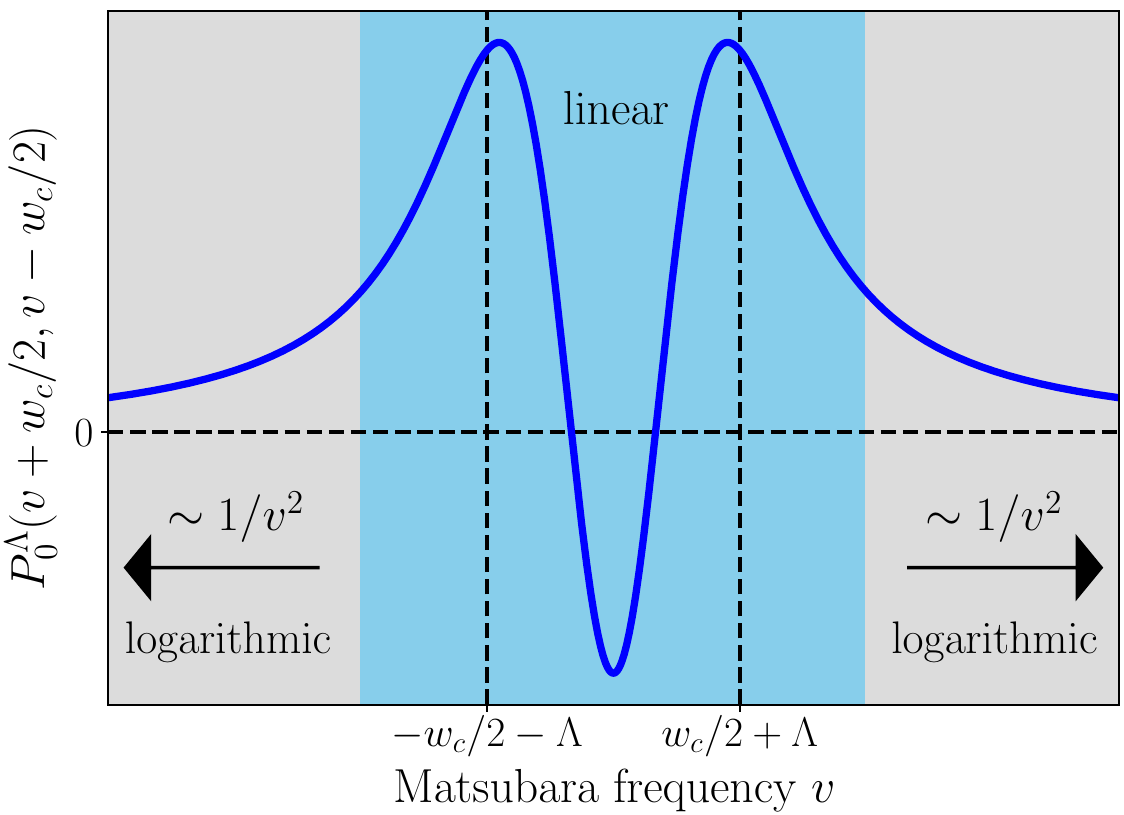}
	\caption{\textbf{Schematic plot of $P^{\Lambda}_{0}$}. The numerical integration of this function can be separated into three domains, each featuring either a multiply peaked structure, where high resolution is required, or a simple $1 / v^2$ decay. While successively lowering the value of the flow parameter, the peaks shift closer to $w_c / 2$ making it necessary to dynamically adjust the breakpoints used for the integration to obtain precise results. See main text for further details on the quadrature rule utilized during the RG flow.}
	\label{fig:bubble}
\end{figure}
In [Fig.~\ref{fig:bubble}] we have schematically plotted the product $P_{0}^{\Lambda}(w, v) = -\big{[} G_{0}^{\Lambda}(w) G_{0}^{\Lambda}(v) \big{]}$ as it typically (up to self-energy corrections) appears for evaluations of the r.h.s of the multiloop flow. By integrating this function one initially generates the frequency dependence of the vertices, and respecting its features is therefore crucial to obtain precise results. \\
The integration domain can roughly be split into three regions, two algebraically decaying tails that enclose a vivid structure residing symmetrically around $v = 0$. 
The position of the peaks is directly related to the transfer frequency of the respective channel $w_c$ as well as the RG scale $\Lambda$ necessitating a dynamical adjustment of integration breakpoints during the flow. Note that, although the outer domains formally require an integration up to infinity, one can in practice cut off the integral at a finite upper bound, where additional contributions to the integral become negligible. \\
To cope with these characteristics we utilize an adaptive quadrature rule, tailored towards the functions at hand. The integration domains are first split into linearly (for the inner domain) or logarithmically (for the outer domains) distributed intervals, where the interval's width is the smallest close to the peaks of $P^{\Lambda}_{0}$ for the logarithmic part. In each of those subdomains, we then apply an adaptive trapezoidal rule ameliorated by a Richardson extrapolation for the final result, where the number of function evaluations is increased until we meet an absolute error tolerance of $10^{-10}$ or a relative error tolerance of $10^{-3}$. \\
The vertices are discretized on non-negative frequency meshes composed of a linear part starting at $w = 0$ with spacing $h$ and a logarithmic part from $Nh$ to some large upper bound, where $N = 0.4 N_{\textrm{tot}}$ is the number of linearly spaced frequencies. Negative frequencies are not used explicitly as they can always be mapped onto their positive counterpart by the symmetries outlined in Appendix \ref{appendix:symmetries}. In total, we monitor seven independent meshes throughout the flow: one for the self-energy ($N_{\Sigma} = 200$) and two for every channel, thereby one for the the transfer frequency axis ($N_{\Omega} = 40$) and one for the fermionic frequency axis ($N_{\nu} = 30$). Decoupling the frequency meshes for the different two-particle channels turned out crucial to stabilize our code for small values of the flow parameter, because competing ground states, paramagnetic ones for the $s/u$ channel and magnetic ones for the $t$ channel, could be resolved in an unbiased way. \\
Finally, the evaluation of the r.h.s. of the flow equations requires knowledge of the vertices for frequencies, which do not necessarily align with the points in the chosen frequency mesh. To address this issue we perform multi-linear interpolations in between grid points for all (at most three) arguments of the different diagram classes, which in the worst case, require eight kernel values to be taken into account.

\subsection{Differential equation solver}

To initialize the RG flow in the ml-FRG framework there are, in principle, two ways. As commonly done in FRG calculations, one can set the initial scale $\Lambda_i$ to a value much larger than the spin coupling $|J| \equiv \sqrt{\sum_i J_i^2}$ (where $J_i$ are the couplings with a finite value in the Hamiltonian) to approximate the $\Lambda \to \infty$ limit where only bare vertices remain. On the other hand, since the ml-FRG converges to the regularized PA by construction, one could also initialize the flow at a somewhat smaller value $\Lambda_i / |J|$ with a solution of the SDE and the BSEs \cite{LMU-Group}. Here, we chose the latter, as it allows to remedy small numerical artifacts, primarily in the self-energy, that appear when the conventional option is selected. \\
Starting from an initial scale $\Lambda_i / |J| = 5$, we therefore first solve the parquet equations by simple fixed point iterations with a damping factor $\beta$ (where $\beta = 1$ corresponds to a full update). The self-energy and two-particle channels are declared to have converged sufficiently once the maximum absolute/relative deviation between two iterations is smaller than $10^{-10}$/$10^{-5}$. In practice, we found quick convergence as long as $\Lambda / |J| > 1$, where no damping was needed to reach the fixed-point, while slowing down rapidly when $\Lambda / |J| \ll 1$. In the latter case smaller and smaller values of $\beta$ were required and directly solving the parquet equations soon became unfeasible, in agreement with \cite{LMU-Group}. \\
The ml-FRG flow equations are integrated using the Bogacki-Shampine method \cite{Bogacki} with adaptive step size control. This causes the flow to first progress rapidly, while slowing down when instabilities, signaling spontaneous symmetry breaking, emerge at smaller energy scales. A third order solver, although it requires multiple (costly) evaluations of the r.h.s of the flow equations, in our opinion resembles a good compromise between reliability and numerical efficiency. We have set an absolute error tolerance of $10^{-10}$ and a relative error tolerance of $10^{-3}$ for one step of the solver, with a minimum step size of $h_{\textrm{min}} = 10^{-4} |J|$ and maximum size $h_{\textrm{max}} = 0.1 \Lambda$, where $\Lambda$ is the current cutoff value in units of $|J|$. To prevent that the step size $h$ increases too rapidly whenever we meet the desired tolerances (and potential features in the flow are therefore overlooked), we limit its growth to at most ten percent with respect to the old value. The RG flow is continued down to a minimal value $\Lambda_f / |J| = 0.05$ if the following sanity checks are fulfilled:
\begin{enumerate}
    \item The absolute maximum of the vertex is smaller than $50 |J|$.
    \item The correlations do not show non-monotonicities like peaks or cusps.
    \item The relative integration error of the ODE solver does not exceed the error tolerance by more than an order of magnitude.
\end{enumerate}
The first and second criterium ensure that the solver is terminated whenever the flow breaks down at some large value of $\Lambda / |J|$ and the step size of the Bogacki-Shampine method therefore diminishes to $h_{\textrm{min}}$, resulting in a critical loss of performance. The last check secures that the adaptive step size control of our ODE solver is still reliable and that $h$ is properly reduced in critical regions of the flow to keep the errors inside the desired bounds. We found the latter test to be occasionally violated when either $\chi^{\Lambda}$ diverges or sufficient convergence in loops cannot be achieved beyond a symmetry-breaking phase transition [see e.g. Fig.~\ref{fig:res_cubic}(a)]. That the flow in these cases becomes unstable is, however, an expected result and the ODE solver is only stopped to prevent excessive run times.
\\
Furthermore, we found that in order to obtain stable results also at small $\Lambda / |J|$, resolving all relevant features of the vertices at different stages of the flow is of special importance. Therefore we have developed a simple scanning routine (cf. Appendix \ref{appendix:scanning}) which analyses the vertices and subsequently proposes a new linear step width for the different frequency meshes after each Runge-Kutta step. The vertices are then transferred to the updated meshes via multi-linear interpolations.

\subsection{Algorithmic complexity}

The asymptotic scaling of computation times with the different numerical parameters can be read off directly from the flow equations and is given by
\[
	\mathcal{O}(N_L^2 \times N_I N_{\Omega} N_\nu^2 \times \ell) \,,
\]
where $N_L \sim L^d$ is the number of symmetry reduced lattice sites for a lattice of dimension $d$, $N_I$ the initial number of linearly/logarithmically spaced intervals for the adaptive frequency integration, $N_{\Omega}$ the number of mesh points for the transfer frequency axis of the channels, $N_{\nu}$ the respective number of points on the fermionic axes and $\ell$ the number of loops. \\
Let us examine in more detail how this scaling is obtained. To do so, we can focus on the computation of the two-particle vertex, as the effort of computing the self-energy derivative, the latter being a function of one frequency argument only, is negligible.
After exploiting lattice symmetries and time translation invariance, each channel is parametrized by one site index, one transfer and two fermionic frequencies. To compute the derivative for each of these components one needs to evaluate the respective right hand side of the flow equations, which comprise a single frequency integration over at least $N_I$ frequency points and, in the case of the $dph$ channel, another summation over the full lattice. Although, for large $\ell$, the number of terms to compute within each loop stays constant, and as such the numerical effort asymptotically scales as $\mathcal{O}(\ell)$, there is a computational overhead going from $\ell = 1$ to $\ell = 3$. The two-loop contribution consists of two terms, left and right part, which both are as costly to evaluate as the one loop terms. Furthermore, for $\ell \geq 3$, the central part additionally comes on top.

\subsection{Code performance}

Given the computational complexity outlined in the previous section, the question arises, how the ml-FRG flow equations can be efficiently integrated down to small values of the infrared cutoff $\Lambda / |J|$, as their number $N_{\textrm{eq}}$ rapidly grows for larger system sizes and increased frequency resolution (in this manuscript for example $N_{\textrm{eq}} \approx 10^7$). Efficient code is therefore crucial to obtain results with modest computational resources and feasible run times. \\
Our code is written in the Julia programming language and so far utilizes two levels of parallelization \cite{Rohe}: vectorization utilizing on-core SIMD units and the invocation of multiple cores per CPU via Julia's native multithreading support. \\
To accelerate the evaluation of the integrands on the r.h.s. of the flow equations, we buffer all spatial contributions for a given tuple of outer frequencies $(w_c, v_c, v_c')$ in an array which is subsequently passed to the adaptive quadrature routine. This not only allows us to recycle interpolation parameters for different lattice sites, but also makes it possible to vectorize the actual read-out process for the vertices. \\
Since different frequency components of the vertex can be computed independently, parallelizing the pf-FRG flow over several cores is in principle straightforward. The largest pitfall in distributing the calculations over multiple threads comes, however, from the adaptiveness of the quadrature routine. This is because every frequency component $(w_c, v_c, v_c')$ may require a different number of integrand evaluations (and therefore computing time) before the trapezoidal rule converges in each domain. In consequence, the workload is highly asymmetric and load balancing becomes vital for boosting code performance to its full extent. The Julia language offers dynamic thread scheduling out of the box and is therefore well suited for this problem. \\
Another possible level of parallelization that could in principle be exploited is the distribution of calculations across multiple computing nodes (for example via MPI). We found, however, that computing times are still tolerable when only a single node is used. For example, a $\ell = 4$ flow for the pyrochlore lattice with $\sim$ 460 sites was obtained in $\sim$ 10 hours with 48 threads on two Intel Xeon Platinum 8168 CPUs. Therefore, distributed memory parallelization is currently not implemented in our code.

\section{Benchmark calculations}

In this section, we present benchmark calculations of our multiloop pf-FRG machinery for a number of (frustrated) quantum spin models -- the Heisenberg antiferromagnet on the pyrochlore lattice, a $J_1 - J_3$ Heisenberg model on the simple cubic (sc) lattice and a $J_1 - J_2$ model on the face-centered cubic (fcc) lattice, with respective Hamiltonians
\begin{align}
    \mathcal{H}_{\mathrm{pyro}} &= J_1 \sum_{\langle ij \rangle} \mathbf{S}_i \cdot\mathbf{S}_j \\
	\mathcal{H}_{\mathrm{sc}}   &= J_1 \sum_{\langle ij \rangle} \mathbf{S}_i \cdot\mathbf{S}_j + J_3      \sum_{\langle \langle \langle ij \rangle \rangle \rangle} \mathbf{S}_i \cdot\mathbf{S}_j \\
	\mathcal{H}_{\mathrm{fcc}}  &= J_1 \sum_{\langle ij \rangle} \mathbf{S}_i \cdot\mathbf{S}_j + J_2 \sum_{\langle \langle ij \rangle \rangle} \mathbf{S}_i \cdot\mathbf{S}_j \,,
	\label{eqn: models}
\end{align}
where the nearest-neighbor coupling $J_1 > 0$ is always antiferromagnetic. Here $J_n$ denotes the spin coupling to the $n$-th nearest neighbor determined by spatial distance. We start by considering two limiting examples, the nearest-neighbor antiferromagnets on the pyrochlore and cubic lattices, respectively. While the former hosts an extensively degenerate (classical) ground state manifold at $T = 0$ and in its quantum version is considered a candidate model for a quantum spin liquid ground state, the latter is free from geometric frustration and features a symmetry-broken ground state at low temperatures \cite{SandvikCubic}, even in the presence of a third-nearest neighbor coupling $J_3$. 
As a final benchmark we consider the phase diagram of the $J_1 - J_2$ model on the fcc lattice, which in its classical limit is interesting for its appearance of degenerate ground state manifolds of codimension $2$ (lines) and $1$ (surfaces) at $T = 0$ \cite{Haar-1962,finnish_93}, thus providing a promising playground to realize a competition between magnetically ordered and quantum spin liquid ground states. 

From a technical point of view, these benchmark calculations show how the multiloop framework can capture the sometimes delicate balance between quantum fluctuations and ordering tendencies. Our case studies provide examples where either one of the two tendencies is strengthened when going to higher loop orders in our pf-FRG calculations.

\subsection{Heisenberg model on the pyrochlore lattice}

\begin{figure*}
    \centering
	\includegraphics[width = 1.0\linewidth]{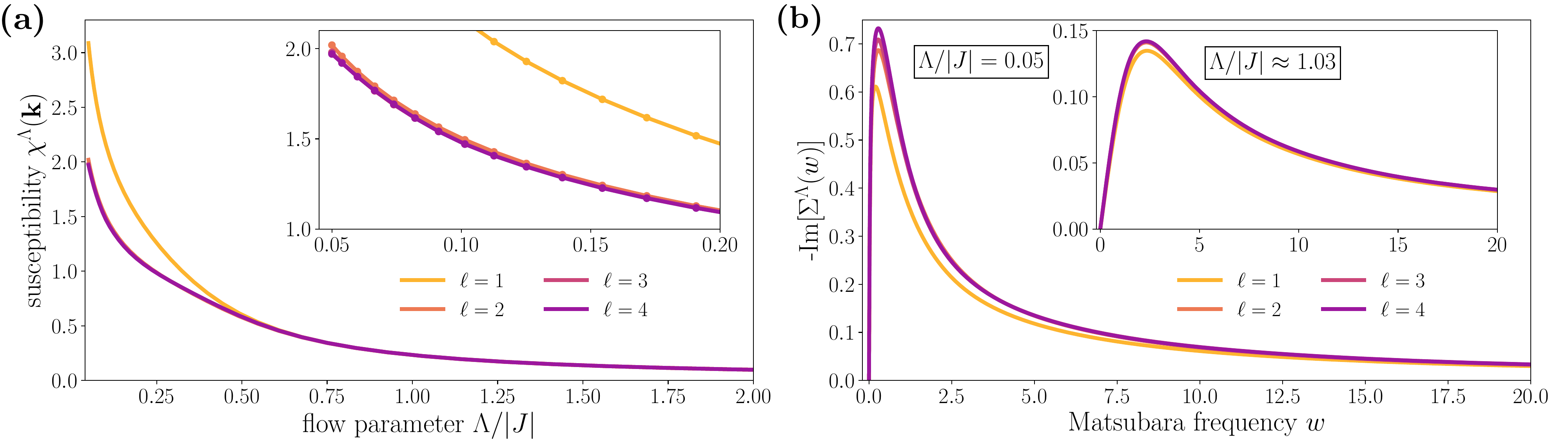}
	\caption{\textbf{Results for the $S = 1/2$ nearest-neighbor Heisenberg antiferromagnet on the pyrochlore lattice} indicating a potential quantum spin liquid ground state. (a) Susceptibility flows depicted at the momentum with the largest amplitude. Increasing the loop order from $\ell = 1$ to $\ell = 4$ leads to rapid convergence (as demonstrated in the inset) and a substantial reduction of $\chi^{\Lambda}(\mathbf{k})$. (b) Multiloop self energies obtained at two different stages of the flow. The inset shows, that two-loop corrections already become relevant at relatively large scales $\Lambda \sim |J|$, with excellent convergence for $\ell > 2$. At small cutoffs, deviations between one and higher loops become more pronounced with respect to position and height of the quasiparticle peak. Though the self energies seem well converged in loops for most frequencies, small differences around the peak are visible, indicating that loop convergence for small $\Lambda / |J|$ on the level of vertices is more difficult to reach than for the spin-spin correlations, in agreement with \cite{LMU-Group}.}
	\label{fig:flows_pyrochlore}
\end{figure*}

\begin{figure*}
    \centering
	\includegraphics[width = 1.0\linewidth]{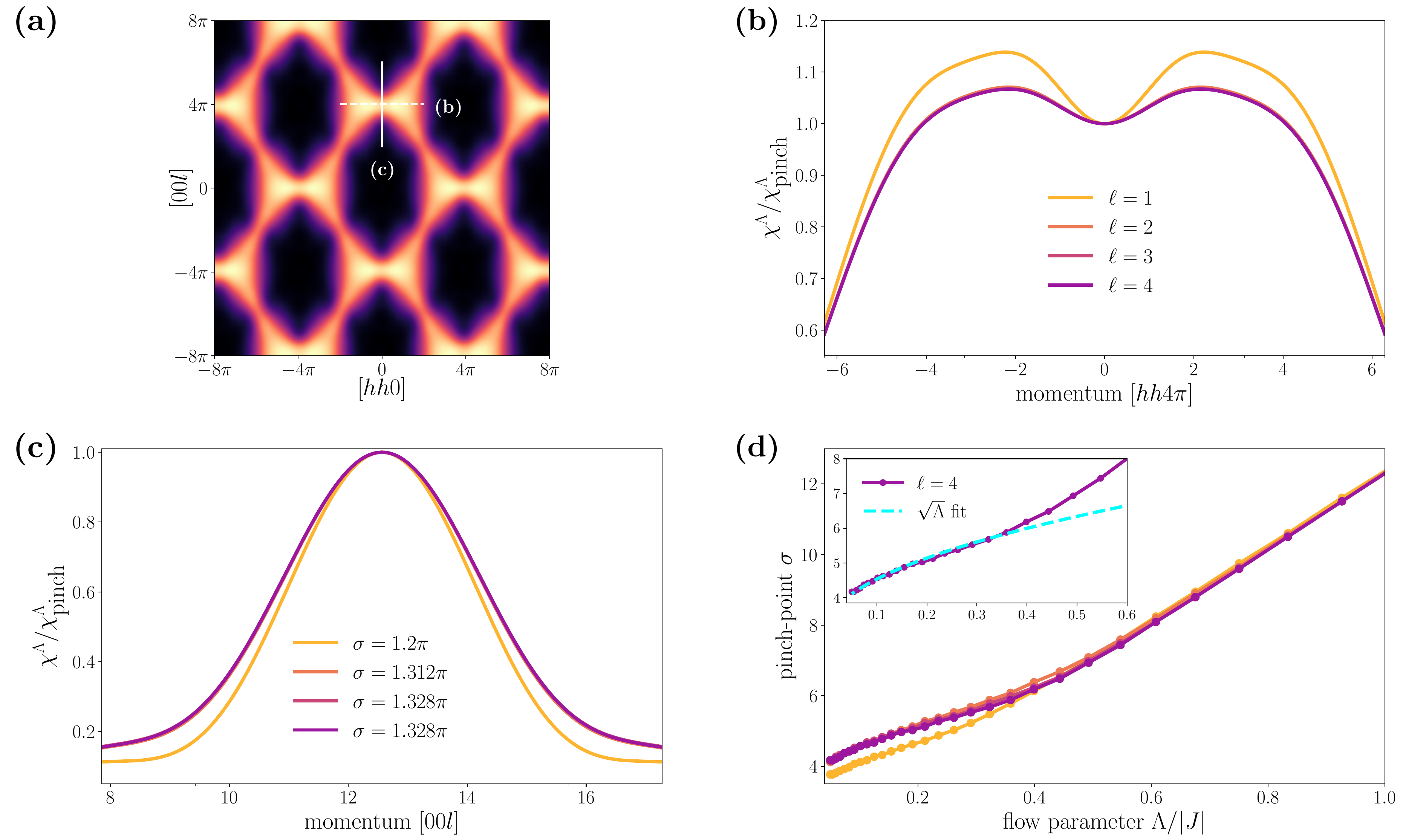}
	\caption{\textbf{Analysis of pinch-points in the momentum resolved susceptibility profile of the nearest-neighbor Heisenberg antiferromagnet on the pyrochlore lattice} at $\Lambda = 0.05 |J|$. (a) Susceptibility in the $[hhl]$ plane for $\ell = 4$. (b) Cut through momentum space along the $[hh4\pi]$ direction as indicated by the dashed horizontal line in (a). (c) Cut through momentum space along the $[00l]$ direction as indicated by the solid vertical line in (a) with the respective full-width-at-half-maximum $\sigma$. The latter increases upon the inclusion of higher loops, in contrast to the classical result, where one expects that the peaks become singular. (d) Flow of $\sigma$ for different loop orders. The inset shows $\sigma$ (for $\ell = 4$) at small values of $\Lambda / |J|$, which to good accuracy obeys a $\sqrt{\Lambda}$ behavior, a result hitherto expected only for the classical model~\cite{Isakov-2004}. However, for large cutoffs $\sigma$ rather scales linear in $\Lambda$.}
	\label{fig:cuts_pyrochlore}
\end{figure*}

The $S = 1/2$ nearest-neighbor Heisenberg antiferromagnet on non-bipartite lattices, such as the kagome or pyrochlore lattices of corner-sharing triangles or tetrahedra, remains an unresolved problem in frustrated quantum magnetism. For the pyrochlore antiferromagnet there is mounting evidence that it hosts a quantum spin liquid ground state~\cite{Canals-1998,Canals-2000,Huang-2016,Chandra-2018,MuellerPyrochlore,Muller-2019}; however, deciphering its nature has proven to be notoriously difficult~\cite{Imre-2020,Astrakhantsev-2021,Derzhko-2020,Schafer-2020,Liu-2019,Kim-2008,Burnell-2009}. In a recent $\ell = 1$ pf-FRG calculation, it was shown that the RG flow of the susceptibility does not develop a divergence at finite $\Lambda$ for any wave vector in the extended Brillouin zone, indicating quantum paramagnetic behavior \cite{MuellerPyrochlore}. Here, we show that this finding is remarkably robust up to $\ell = 4$, where our results have sufficiently converged [see Fig.~\ref{fig:flows_pyrochlore}], providing compelling evidence in favor of a quantum paramagnetic ground state. This low-temperature  phase is characterized by the presence of a bowtie pattern in the susceptibility profile of the $[hhl]$, i.e., $k_{x} = k_{y}$ plane~\cite{Isakov-2004}, with the points at the centre of the bowties (called ``pinch points'') being host to sharp features (singularities) in the case of the corresponding classical model at $T = 0$~\cite{Zinkin-1997}. These pinch points are reflective of dipolar spin correlations~\cite{Huse-2003,Moessner-2003} which are hallmark of a cooperative paramagnetic state -- a Coulomb phase \cite{Henley2010coulomb}, and have been argued to arise from the zero total spin moment rule (called ice-rule) on every tetrahedron~\cite{Chalker-1998,Moessner-1998,Isakov-2004}. In contrast, for a quantum model it is impossible to have a vanishing magnetization on every tetrahedron because the Hamiltonian does not commute with the total spin operator of any given tetrahedron. Hence, quantum fluctuations lead to violations of the ice-rule, with the pinch points losing their sharpness and their singularity rounded off. Consequently, the pinch points [the $(0, 0, \pm 4 \pi)$ (and symmetry related) points in Fig.~\ref{fig:cuts_pyrochlore}(a)] smear out, acquiring a finite-width~\cite{Schafer-2020,Zhang-2019,Plumb-2019,MuellerPyrochlore,Muller-2019,Huang-2016,Canals-1998,Harris-1991}, which serves as a measure of the degree of violation of the ice-rule, i.e., the net magnetization acquired by the tetrahedra. In order to get a quantitative picture concerning the impact of diagrammatic contributions at higher loop orders, we plot the susceptibility along the $h=0$ momentum cut [the vertical solid white line in Fig.~\ref{fig:cuts_pyrochlore}(a)] for different loop orders in [Fig.~\ref{fig:cuts_pyrochlore}(c)]. One observes that the width of the pinch point, as quantified by the full-width-at-half-maximum ($\sigma$) {\em increases} with $\ell$ and finally converges at $\ell=4$ to $\sigma=1.328\pi$ compared to $\sigma=1.2\pi$ at $\ell=1$ [see Refs.~\cite{MuellerPyrochlore,Muller-2019,Schafer-2020} for comparison of $\sigma$ with other methods]. This finding suggests that in the pyrochlore Heisenberg antiferromagnet {\em quantum fluctuations get amplified with increasing loop order}. In Fig.~\ref{fig:cuts_pyrochlore}(d), we show the evolution of $\sigma$ with $\Lambda$ (effective temperature)~\cite{Iqbal3D}, and find that it remarkably obeys (to a good accuracy) the same $\sqrt{\Lambda}$ scaling at small $\Lambda$ expected of a classical model~\cite{Isakov-2004}. \\
The variation of the intensity along a horizontal cut through the pinch point [dashed horizontal line in Fig.~\ref{fig:cuts_pyrochlore}(a)] is shown in Fig.~\ref{fig:cuts_pyrochlore}(b). It is interesting to note that the maxima of the static susceptibility in the $[hhl]$ plane is not located at the pinch points [$(0,0,\pm4\pi)$] but rather in the two symmetrical lobes of the bowties in agreement with Ref.~\cite{Canals-1998}. This should be compared with the findings from a recent finite-temperature matrix product state study~\cite{Schafer-2020,Imre-2020} on clusters up to 128 sites (with fully periodic boundary conditions) which located the maxima of the equal-time structure factor $S({\mathbf q})$ at the pinch points. Given the fact that all but two of the cluster geometries considered in this study do not preserve the full cubic pyrochlore symmetry, it is difficult to reliably establish the behavior of $S({\mathbf q})$ in the thermodynamic limit. A rotation-invariant Green’s function method (RGM)~\cite{Muller-2019} (computing $S({\mathbf q})$) and bold-diagrammatic Monte Carlo simulations (computing static susceptibility)~\cite{Huang-2016} find the intensity distribution to be essentially constant across the length of the bowtie. This variance in the findings between the three methods calls for further investigations since these different patterns of intensity distributions likely correspond to different quantum spin liquid mean-field ans\"atze~\cite{Liu-2019}. Hence, for an accurate identification of the nature of the quantum spin liquid ground state of the $S = 1/2$ Heisenberg antiferromagnet on the pyrochlore lattice, which still remains at large, it will be important to unambiguously resolve the behavior of $S({\mathbf q})$ and the static susceptibility in the thermodynamic limit from other numerical approaches. 

From a purely methodological perspective, we have demonstrated that loop convergence for the pyrochlore antiferromagnet can be obtained even at small values of the cutoff (percent level relative to the bare coupling) and already with a modest number of loops ($\ell \approx 4$). 

\subsection{$J_1 - J_3$ Heisenberg model on the cubic lattice}

\begin{figure*}[t]
    \centering
	\includegraphics[width = 1.0\linewidth]{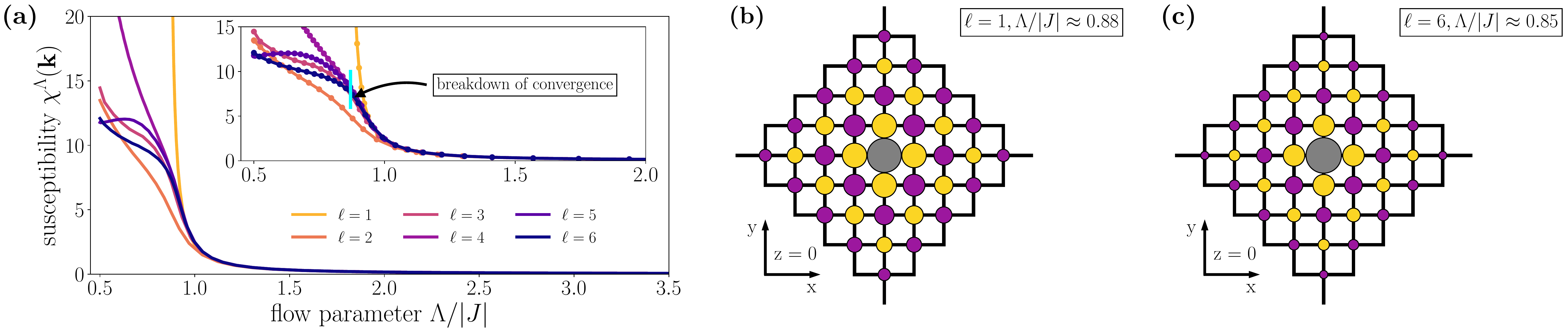}
	\caption{\textbf{Results for the $S = 1/2$ nearest-neighbor Heisenberg antiferromagnet on the simple cubic lattice}. (a) Susceptibility flows at the dominant momentum $\mathbf{k} = (\pi, \pi, \pi$). With increasing loop order the $\ell = 1$ divergence is rounded off to a gentle shoulder in the $\ell > 1$ flows. However, beyond $\Lambda / |J| \approx 0.85$ (marked by the vertical turquoise line in the inset) the multiloop flows cannot be properly converged indicating a breakdown of ml-FRG and therefore a phase transition. (b) Normalized real space correlations in the $z = 0$ plane for a $L = 4$ patch of the full lattice obtained from $\ell = 1$ calculations right before the divergence. Here, purple (yellow) dots denote positive (negative) values of $\chi_{i_0 j}$ where the reference site $i_0$ is marked by a grey circle. (d) Same as (c) but for $\ell = 6$ at the point where loop convergence breaks down.}
	\label{fig:res_cubic}
\end{figure*}

We now turn our attention to a similar Heisenberg type Hamiltonian, but for a lattice geometry devoid of any geometric frustration -- the simple cubic lattice, which we however augment by a third nearest-neighbor interaction $J_3$. This model system exhibits a magnetically ordered ground state for all couplings, with a transition from staggered Ne\'el to collinear magnetic order for ferromagnetic $J_3 < -0.3~J_1$.
Indeed, quantum Monte Carlo simulations \cite{SandvikCubic, Iqbal3D} have confirmed that the model orders at relatively large temperatures  $T_c / |J| \sim 1$, a result which could already be reproduced by previous one-loop pf-FRG calculations \cite{Iqbal3D}. For $J_3 < 0$, however, exchange frustration sets in and QMC approaches are not applicable due to the negative sign problem, though the classical ground states (at $T = 0$) are non-degenerate and the magnetic order simply changes from staggered to collinear at $J_3 / J_1 = -0.25$. Here, we probe the effect of quantum fluctuations on the phase transition in the frustrated regime. 

\begin{figure}
    \centering
	\includegraphics[width = 1.0\linewidth]{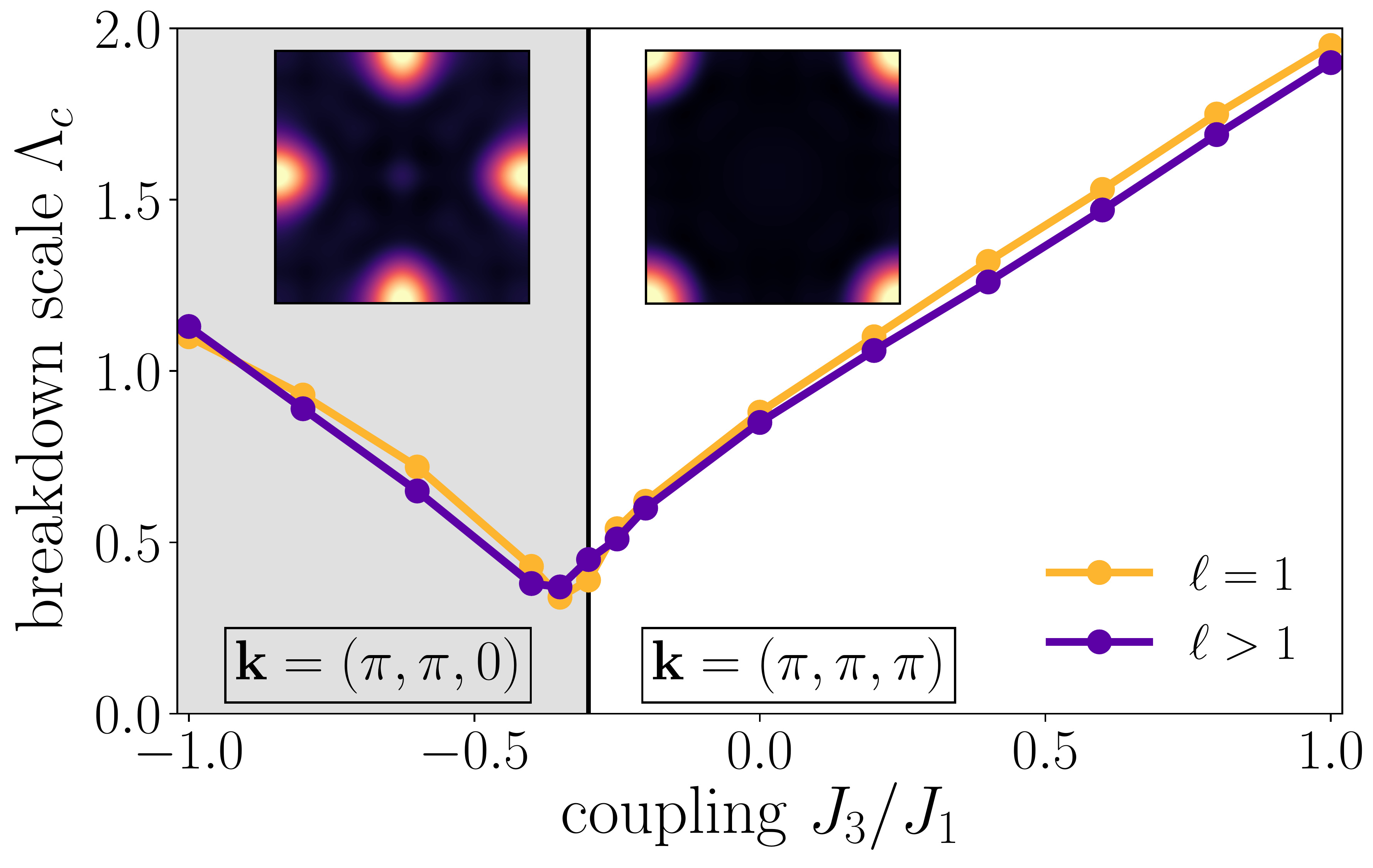}
    \caption{\textbf{Phase diagram for the $J_1 - J_3$ model on the simple cubic lattice}. The pf-FRG data is obtained from one and higher loop calculations with atmost $\ell = 6$. Antiferromagnetic third-nearest neighbor couplings $J_3 > 0$ stabilize Ne\'el order with wave vector $\mathbf{k} = (\pi, \pi, \pi)$. Ferromagnetic $J_3 < 0$ introduces exchange frustration, leading to a suppression of the breakdown scale $\Lambda_c$ for intermediate values $-0.4 < J_3 / J_1< 0.0$. At $J_3 / J_1 = -0.3$ (vertical black line), the magnetic order changes, promoting the momentum $\mathbf{k} = (\pi, \pi, 0)$ instead. Insets show the static susceptibilities (for $\ell = 6$) in the two phases, plotted in the first Brillouin zone for $k_z = \pi$. Though $\Lambda_c$ slightly deviates between one and higher loops, the results are qualitatively consistent.}
	\label{fig:phase_diagram_cubic}
\end{figure}

To start our analysis, we consider the limit $J_3 = 0$ and study the impact of higher loops on the formation of magnetic order for the cubic antiferromagnet [see Fig.~\ref{fig:res_cubic}(a)]. On the one loop level, the susceptibility flow diverges at $\Lambda_c / |J| \approx 0.86$, where the real-space correlations are in line with an antiferromagnetic ground state [Fig.~\ref{fig:res_cubic}(b)], consistent with Refs.~\cite{SandvikCubic, Iqbal3D}. When higher loop orders are included, the one loop divergence is diminished and only a soft shoulder appears in the $\ell > 1$ flows, though in close vicinity to the former. In addition, we were not able to properly converge the multiloop flows beyond $\Lambda_c / |J| \approx 0.85$ [see the inset in Fig.~\ref{fig:res_cubic}(a)] with the errors produced by our Runge-Kutta method growing relatively large such that the step size of the ODE solver was drastically reduced. This indicates that, although a true divergence seems difficult to accurately resolve at higher loops, the description of the physical ground state by ml-FRG, which has the full symmetries encoded by construction, breaks down, signaling a transition into a symmetry broken phase.
Considering the rapid convergence at higher loops for the pyrochlore model even at an order of magnitude smaller values of the cutoff, we can exclude, that the non-systematic behavior we observe beyond $\Lambda / |J| \approx 0.85$ in the present case, is due to the numerical stability of our implementation. The correlations computed for $\Lambda_c / |J| \gtrsim 0.85$, where our flows still converged sufficiently well, support the formation of antiferromagnetic order [Fig.~\ref{fig:res_cubic}(c)], though their range and amplitude are reduced with respect to the $\ell = 1$ result [compare Fig.~\ref{fig:res_cubic}(b) \& (c)]. \\
For finite $J_3$ we found the behavior between one and higher loops to qualitatively agree with our findings for the nearest-neighbor antiferromagnet. Using the absence of loop convergence as an indicator for breakdown of the ml-flow, we coarsely scanned the phase diagram of the $J_1 - J_3$ model in the frustrated ($J_3 < 0$) and non-frustrated ($J_3 > 0$) parameter regime [see Fig.~\ref{fig:phase_diagram_cubic}], determining the critical scales $\Lambda_c$ where the flow cannot be faithfully continued for $\ell = 1$ and $\ell > 1$. We find that both at the one and higher loop level, $\Lambda_c$ is suppressed close to the phase transition from $(\pi, \pi, \pi)$ to $(\pi, \pi, 0)$ magnetic order at $J_3 / J_1 = -0.3$. The value of the breakdown scale is similar between $\ell = 1$ and $\ell > 1$, with the higher loop result being slightly smaller in most cases. Given that the classical phase boundary at $J_3 / J_1 = -0.25$ lies in close vicinity to our FRG result, we conclude, that quantum fluctuations, which were boosted for the strongly frustrated pyrochlore model, have only little influence on the ground state of the mildly frustrated model at hand.

\subsection{Heisenberg model on the fcc lattice}

\begin{figure}
    \centering
	\includegraphics[width = 1.0\linewidth]{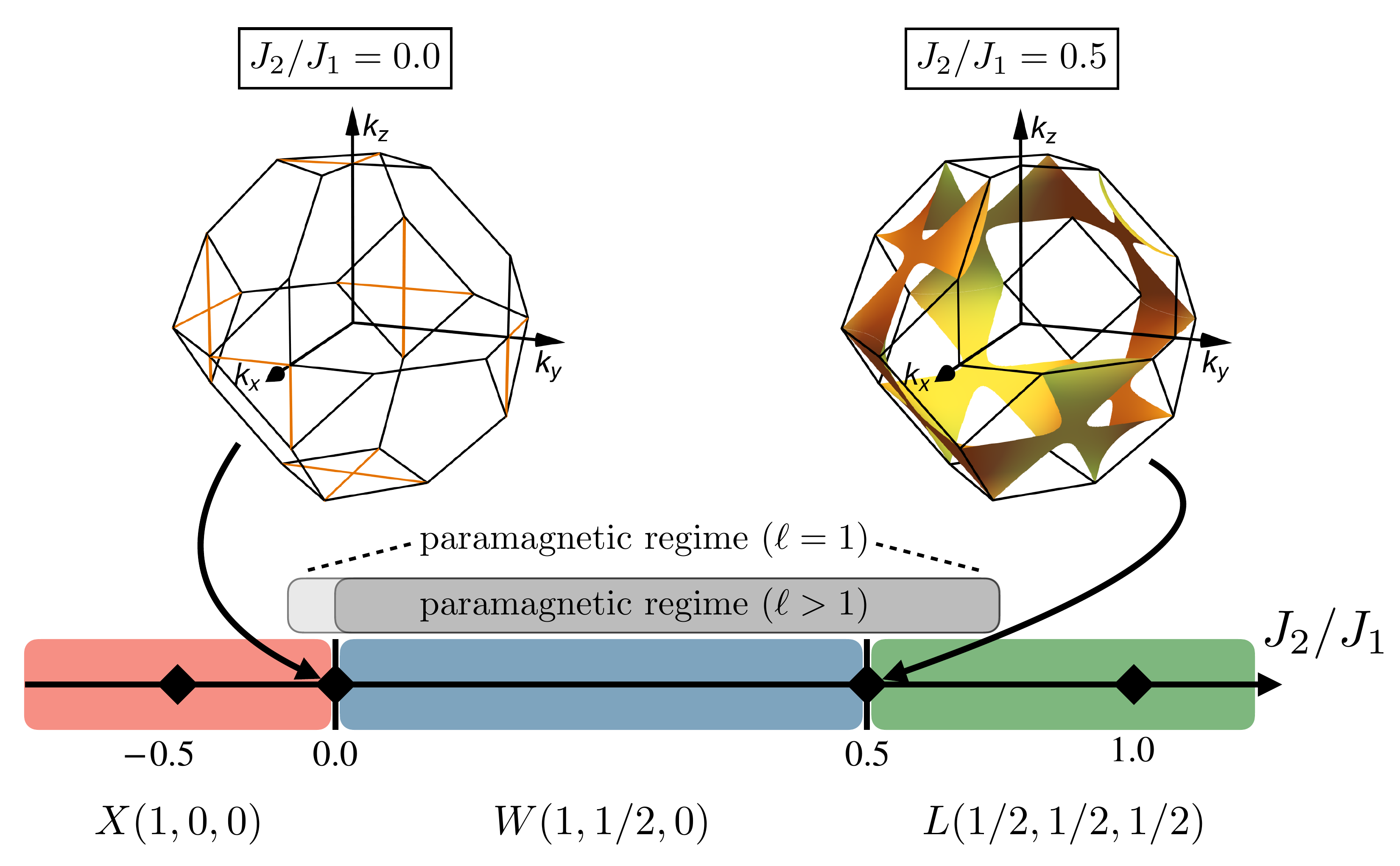}
	\caption{\textbf{Phase diagram of the $J_1-J_2$ fcc model} at $T = 0$. Classically, one finds three phases with energetically favorable momenta located at the $X$, $W$ and $L$ high symmetry points of the first Brillouin zone. These magnetic states are separated by two points, at $J_2 / J_1 = 0.0$ and $J_2 / J_1 = 0.5$, where sub-extensively degenerate ground state manifolds (lines and surfaces) appear. In the quantum model, we find an extended regime ($J_2 / J_1 \approx 0.0 - 0.65$) without a breakdown of the (ml-) FRG flows, marking a possible realm to realize quantum spin liquid behavior. Increasing the loop order leads to a small decrease in the extend of the paramagnetic regime with respect to the lower bound, which is shifted from $J_2 / J_1 \approx -0.1$ for $\ell = 1$ to $J_2 / J_1 \approx 0.0$ for $\ell > 1$. Black markers indicate the couplings for which we display the results more explicitly in Fig.~\ref{fig:res_fcc}.}
	\label{fig:res_fcc_LT}
\end{figure}

The face centered cubic (fcc) crystal structure serves as another classic textbook example of a three-dimensional Bravais lattice which is not bipartite, thereby frustrating the two-sublattice N\'eel order. A measure of the degree of frustration is provided for by the dimensionality of the ground state manifold (GSM), i.e., the set of wave vectors $\{\mathbf Q\}$ where $J(\mathbf{q})$ takes on its minimal value. At $T = 0$, the corresponding classical ($S \to \infty$) version of Eq.~\eqref{eqn: models} with $J_{2} = 0$ features a one-dimensional degenerate GSM \cite{Haar-1962,Alexander-1980} [see left plot in Fig.~\ref{fig:res_fcc_LT}], while at $J_{2} = 0.5$, the GSM takes the form of a two-dimensional spin spiral surface \cite{finnish_93} 
\begin{equation}
    \cos \frac{Q_{x}a}{2}+\cos \frac{Q_{y}a}{2}+\cos\frac{Q_{z}a}{2}=0,\label{surf_eq}
\end{equation}
[see right plot in Fig.~\ref{fig:res_fcc_LT}] reflective of an increased frustration. This two-dimensional manifold can be topologically characterized as a triply periodic Schwarz-P surface with an Euler characteristic $\chi = -4$ \cite{bubble_surfaces}, rationalized by an affine lattice construction \cite{Balla-2019}, and can be associated with an electronic Fermi surface via a supersymmetry construction \cite{Attig2017}. \\
The origin of these degeneracies is manifest once the Hamiltonian is recast as a sum of complete squares of spins over edge-sharing tetrahedra (for $J_{2} = 0$) and edge sharing octahedra (for $J_{2} = 0.5$) which tessellate the fcc lattice \cite{IqbalFCC}
\begin{align}
\mathcal{H}&=\frac{J_1}{4}\sum_{\textrm{tetra}}\left(\mathbf{S}_1+\mathbf{S}_2+\mathbf{S}_3+\mathbf{S}_4\right)^2 - 2J_{1}N \label{eq:nndeg}\\
\mathcal{H}&=\frac{J_1}{4}\sum_{\textrm{octa}}\left(\mathbf{S}_1+\mathbf{S}_2+\mathbf{S}_3+\mathbf{S}_4+\mathbf{S}_5+\mathbf{S}_6\right)^2-\frac{3}{2} J_1 N \label{eq:j2deg},
\end{align}
where $\mathbf{S}_1, \ \dots , \ \mathbf{S}_4$ [Eq.~\eqref{eq:nndeg}] and $\mathbf{S}_1, \ \dots , \ \mathbf{S}_6$ [Eq.~\eqref{eq:j2deg}] refer to the four and six spins on the sites of a tetrahedron and octahedron, respectively. Since $J_{1} > 0$, the Hamiltonian is minimized if and only if the spins sum up to zero on every tetrahedron (for $J_{2} / J_{1} = 0$) and octahedron (for $J_{2} / J_{1} = 0.5$), with the additional constants giving the ground state energy. Every spin configuration satisfying this zero magnetization constraint is a valid classical ground state at $T = 0$. When the temperature $T \neq 0$ or/and the reciprocal spin $1/S \neq 0$, thermal and quantum fluctuations could potentially lift this degeneracy via the entropic order-by-disorder mechanism~\cite{Villain-1980} and stabilize long-range magnetic order. However, if they fail to do so, one realizes a quantum paramagnet which could possibly be a quantum spin liquid. Thus, the $J_{1}$\textendash$J_{2}$ fcc Heisenberg antiferromagnet serves as an ideal testbed to study the role of diagrammatic contributions at higher loop orders in distilling the nontrivial and subtle interplay of quantum and thermal selection effects for $S = 1/2$. \\
\begin{figure*}
    \centering
	\includegraphics[width = 1.0\linewidth]{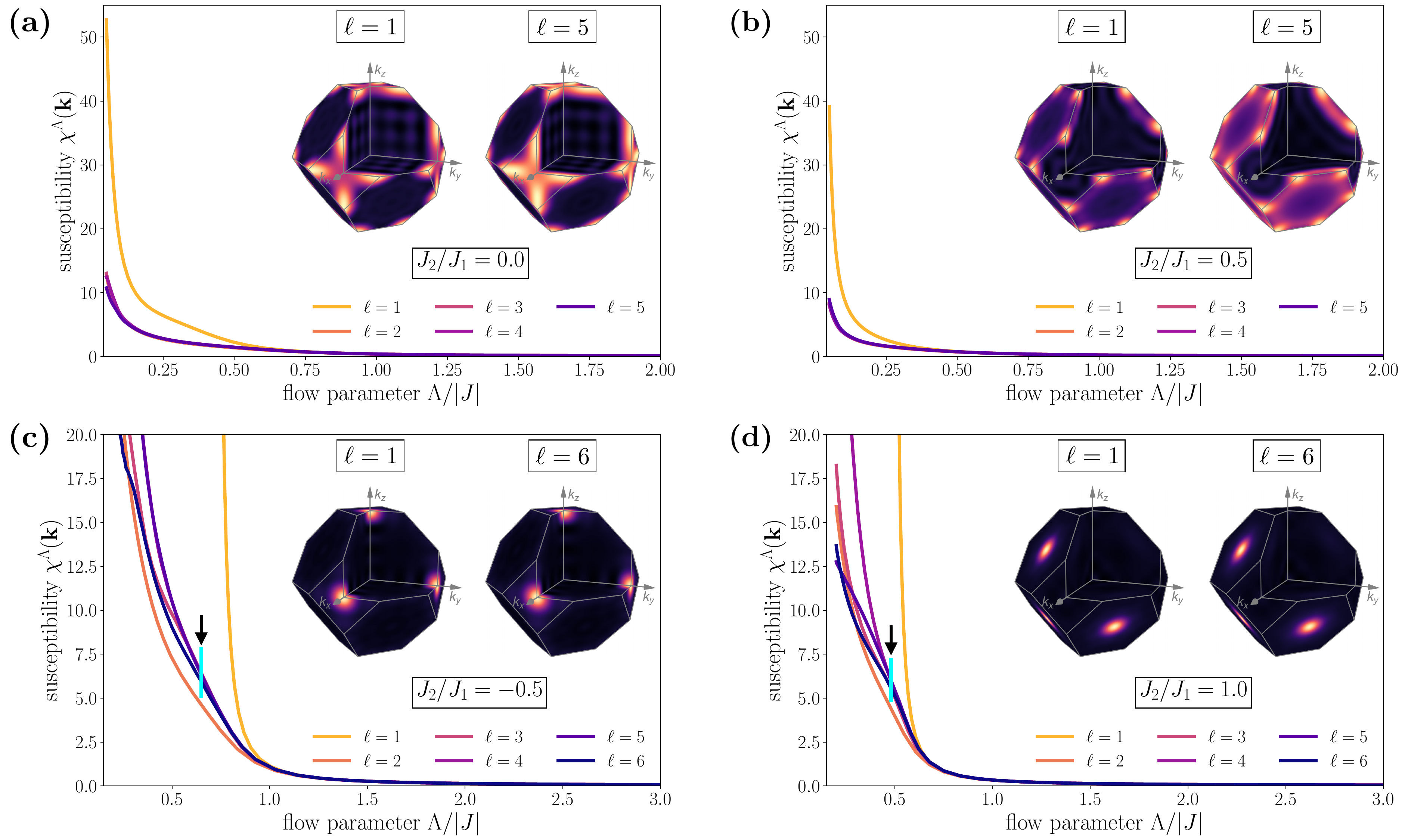}
	\caption{\textbf{Multiloop results for the $J_1-J_2$ Heisenberg model on the fcc lattice.} (a) Susceptibility flows for the fcc antiferromagnet, showing smooth and converging (for $\ell \leq 4$) flows down to the numerical lower bound set for $\Lambda$. Insets show the momentum resolved susceptibilities in the first Brillouin zone, extracted at the lowest possible cutoff. (b) Same as (a) but for $J_2 / J_1 = 0.5$. (c) \& (d) Results for $J_2 / J_1 = -0.5$ and $J_2 / J_1 = 1.0$ deep in the ordered phases [see Fig.~\ref{fig:res_fcc_LT}]. In contrast to the one loop flow, which diverges at a finite $\Lambda_c$, the multiloop results remain regular, though they could not be converged far below the characteristic scale of the one loop result as indicated by a thick turquoise line. The suceptibilities plotted in the inset are computed from the results right before the divergence for $\ell = 1$ and before loop convergence breaks down for $\ell = 6$.}
	\label{fig:res_fcc}
\end{figure*}
For the nearest-neighbor $S = 1/2$ Heisenberg antiferromagnet [see Fig.~\ref{fig:res_fcc}(a)], we find that the RG flows of the susceptibility at loop orders $\ell \leq 5$ do not display a divergence at finite $\Lambda / |J|$ for the wave vectors of either of the two classically degenerate orders present for $J_{2} = 0$, namely, the $X(1, 0, 0)$ [Type-I] or $W(1, 1/2, 0)$ [Type-III] orders (we henceforth adopt the notation where the points in the Brillouin zone are referred to by their names and coordinates in units of $2 \pi$, e.g., $Q_{W}=(2 \pi, \pi, 0) = W(1, 1/2, 0)$, where the lattice constant $a = 1$). Furthermore, one observes that the susceptibility displays strongly broadened maxima at the $W(1, 1/2, 0)$-points [see $\ell = 5$ in Fig.~\ref{fig:res_fcc}(a)], consistent with earlier pf-FRG calculations \cite{KieseFCC}, resembling the classical lines of degeneracy. Although loop convergence is excellent up to $\ell = 4$, small deviations become visible for $\ell = 5$ at the smallest cutoffs. We attribute the latter to numerical interpolation errors which become stronger the higher the loop order and the lower $\Lambda$, rather than a breakdown of the flow (as e.g. in the cubic antiferromagnet), since lower loop orders were shown to be converged already. Further simulations with even higher frequency resolution will presumably remedy these artifacts and allow flows with large $\ell$ towards and beyond the minimum cutoff value of $\Lambda / |J| = 0.05$ chosen here. This is likely to shed light towards addressing a long-standing problem of whether the ground state of the $S=1/2$ fcc Heisenberg antiferromagnet develops long-range magnetic order or is nonmagnetic in nature. The latter scenario (for which we see some signatures) provides a rare example of a frustrated model with a codimension-2 manifold where the combined effect of quantum and thermal fluctuations fails to lift the degeneracy thus realizing a paramagnetic ground state. \\
The classically degenerate point $J_{2}/J_{1} = 0.5$ is a triple point of the  $W(1, 1/2, 0)$ [Type-III], $L(1/2, 1/2, 1/2)$ [Type-II], and incommensurate spiral $(q, q, 0)$ orders \cite{Yamamoto-1972}, and the RG flows for the susceptibility tracked at the corresponding wave vectors display a smooth and monotonically increasing behavior down to the lowest simulated cutoff at all loop orders, with well converged results to $\ell = 5$ [see Fig.~\ref{fig:res_fcc}(b)], similar to the non-magnetic ground state probed for the pyrochlore antiferromagnet. The absence of a divergence at finite-$\Lambda$ (and the rapid convergence for $\ell > 1$) provides strong evidence in favor of a quantum paramagnetic ground state \footnote{We cannot rule out a scenario where long-range magnetic order sets in at $\Lambda / |J| < 0.05$, as has been hinted to in Ref.~\cite{Sun-2018}}~\cite{Ignatenko-2008}. With increasing loop order one observes a progressive smearing and softening of the spectral weight [compare $\ell = 1$ and $\ell = 5$ in Fig.~\ref{fig:res_fcc}(b)], and at $\ell = 5$ order we have a broadly homogeneous distribution of intensity over the surface of the Brillouin zone with soft maxima at the $W(1, 1/2, 0)$ points. A recent work \cite{Sonnenschein-2020} has identified two symmetric $\mathds{Z}_{2}$ quantum spin liquids which could potentially serve as candidate ground states, (i) a gapped $\mathds{Z}_{2}$ state and (ii) a $\mathds{Z}_{2}$ spin liquid featuring a network of symmetry-protected linelike zero modes in reciprocal space. Within a self-consistent mean-field treatment the state (ii) was found to have a lower energy with the corresponding dynamical spin structure factor exhibiting enhanced intensity at the $L(1/2, 1/2, 1/2)$ point. This finding lends support to a scenario whereby a redistribution of spectral weight from the $W(1, 1/2, 0)$ to the $L(1/2, 1/2, 1/2)$ is likely to occur at a relatively lower energy scale. In contrast to $S = 1/2$, in the semiclassical limit $(1/S \ll 1)$, quantum fluctuations (treated within the harmonic approximation) have been shown to select the $L(1/2, 1/2, 1/2)$ [Type-II] long-range magnetically ordered state~\cite{Balla-2019}. \\
For $J_{2} / J_{1} = -0.5$ ($J_{2} / J_{1} = 1.0$), i.e., deep in the magnetically ordered phases of the classical model, our results display the same behavior that we have observed for the symmetry broken ground state of the cubic lattice antiferromagnet [see Fig.~\ref{fig:res_fcc}(c) and (d)]. At the one loop level, the susceptibility flows at the $X(1, 0, 0)$ (or $L(1/2, 1/2, 1/2)$ respectively) points diverge, with clearly resolved incipient Bragg peaks in the corresponding momentum resolved susceptibilities. While the structure factors do not change qualitatively at higher loops, the divergence vanishes, though for cutoffs close to the respective $\Lambda_c / |J|$ of the $\ell = 1$ flows, loop convergence cannot be achieved up to $\ell = 6$. \\
Finally, we performed a rough scan of the full phase diagram of the antiferromagnetic $J_1-J_2$ fcc model. In between the two degenerate points $J_2 / J_1 = 0.0$ and $J_2 / J_1 = 0.5$, we found an extended regime of paramagnetic ground states [see grey bar in Fig.~\ref{fig:res_fcc_LT}], where one and higher loop results consistently show no flow breakdown. Furthermore, at $J_2 / J_1 \approx 0.65$ both calculations with $\ell = 1$ and $\ell > 1$ predict a transition into the $L(1/2, 1/2, 1/2)$ ordered state. For $J_2 / J_1 \lesssim -0.1$ on the other hand, one loop calculations predict Type-I magnetic order, whereas higher loop calculations show no loop convergence down to the lowest cutoffs for $J_2 / J_1 \lesssim 0.0$. The extend of the putative spin-liquid regime is therefore slightly reduced between $\ell = 1$ and $\ell > 1$.

\section{Conclusions and outlook}

In this manuscript, we set out to add several methodological refinements to the pf-FRG approach for quantum spin models. 
Our primary goal was a transcription of the multiloop truncation scheme \cite{Kugler_1, Kugler_2} to pf-FRG, whose numerical implementation the manuscript at hand describes in meticulous detail. On a technical level, we found that the implementation of the multiloop pf-FRG approach necessitates a critical reevaluation of the (adaptive) integration schemes employed in solving the coupled integro-differential equations, particularly with regard to the underlying frequency discretization. These methodological advancements we make accessible via the open-source package \texttt{PFFRG.jl} written in the Julia programming language~\cite{PFFRGjl}. \\
As a benchmark, we have employed this multiloop pf-FRG approach to a family of Heisenberg antiferromagnets, subject to varying levels of geometric frustration. For the model with the highest degree of frustration, i.e. the pyrochlore antiferromagnet (the GS is extensively degenerate in the classical limit), we find that the multiloop corrections strengthen quantum fluctuations, which we decipher via a careful analysis of the width of the pinch points characterizing the low-temperature quantum spin liquid phase. In addition, we found excellent convergence of $\chi^{\Lambda}$ already for $\ell = 4$ even at the smallest cutoff $\Lambda / |J| = 0.05$. These results are to be contrasted with the data obtained for the cubic antiferromagnet, which, due to the bipartite nature of the underlying lattice, is free from geometric frustration. Though we have shown that resolving a divergence of $\chi^{\Lambda}(\mathbf{k})$ for $\ell > 1$ is rather challenging due to long computation times and large errors of the ODE solver, we could demonstrate that converging the multiloop flows was not possible far beyond the characteristic scale of the one loop result, indicating a breakdown of ml-FRG in this regime. Furthermore, the real space correlations, momentum resolved susceptibilities and phase boundary, when a finite third nearest-neighbor coupling $J_3$ is included, are qualitatively consistent between $\ell = 1$ and $\ell > 1$. These two different scenarios, loop convergence at small cutoffs for putative spin liquids and the absence thereof when $SU(2)$-symmetry is spontaneously broken, were shown to be consistent with our findings for the $J_1-J_2$ Heisenberg model on the fcc lattice, settled between the cubic and pyrochlore antiferromagnets frustrationwise. Enclosed between the two degenerate points $J_2 / J_1 = 0.0$ and $J_2 / J_1 = 0.5$ we found an extended regime of paramagnetic states, whose full extend is, however, slightly reduced when higher loop calculations are employed. \\
Even though the inclusion of higher loop orders for FRG calculations on itinerant fermion systems has demonstrated, that already with a few iterations, convergence in several susceptibilities can be reached \cite{HilleQuantitative, HillePseudogap}, for spin systems as considered here one could not anticipate that the RG flow is similarly well behaved. Formally, since the spinons do not carry kinetic energy, our parton decomposed Hamiltonian resembles the $U \to \infty$ limit of the Hubbard model and consequently there is no small parameter that one can built a perturbative argument on. Remarkably, our work shows that convergence in loop order can be achieved also for an FRG treatment of {\em strongly coupled} pseudo-fermions, complementing the initial development of multiloop FRG in the weakly coupled regime \cite{Kugler_1, Kugler_2}. \\
Employing the multiloop pf-FRG may pave an avenue for further systematic improvements. Besides the demonstration of loop convergence on the level of post-processed susceptibilities, no difference between the latter and susceptibilities computed from response functions should remain at higher loops \cite{Tagliavini}. Similarly, self-energies and two-particle vertices should converge to solutions of the regularized PA at all cutoffs where the symmetries of the microscopic model are preserved. Note that, for moderate cutoffs, this has been shown in Ref.~\cite{LMU-Group}. Furthermore, the generalization of our formulation to Hamiltonians with reduced spin symmetries \cite{PhysRevLett.108.127203,Hering-2017,BuessenOffDiag} or additional degrees of freedom \cite{KieseSpinValley} is in principle straightforward, as it does not alter the principal structure of the multiloop equations. \\
Finally, a longterm objective of the further development of the pseudo-fermion FRG is to gain access not just to static correlators, but also to dynamic correlations of frustrated quantum spin systems in order to facilitate an in-depth comparison between microscopic theoretical modeling and experimental evidence from, e.g., neutron scattering. It is likely that for all such enterprises, the refinements of pf-FRG reported in this work are vital to achieve sufficient numerical performance.

\acknowledgements

While preparing this manuscript we learned of a parallel effort to implement multiloop pf-FRG by a group in Munich consisting of J. Thoenniss, M. K. Ritter, F. B. Kugler, J. von Delft and M. Punk. We cordially thank them for insightful and detailed discussions on the subject, the open exchange of ideas and comments and for coordinating the release of our papers \cite{LMU-Group}. 
We thank J. Reuther for insightful discussions on the technical aspects of pf-FRG and the implementation of higher loop orders therein. \\
D. K. thanks M. M. Scherer, L. Gresista and D. Rohe for fruitful discussion on different aspects of pf-FRG and its numerical implementation. \\
Y.I. thanks P. Balla, F. Becca, F. Ferrari, K. Penc, J. Richter, J. Sonnenschein, and M. Zhitomirsky for helpful discussions and collaboration on related topics. \\
R.T. thanks L. Balents, A. V. Chubukov, M. Gingras, W. Metzner, F. Mila, R. Moessner, N. Perkins, M. Rice, M. Salmhofer, O. Starykh, O. Tchernyshyov, F. Verstraete, and J. Vidal for discussions. \\
The Cologne group acknowledges partial support from the Deutsche Forschungsgemeinschaft (DFG) -- Projektnummer 277146847 -- SFB 1238 (project C02). \\
The W\"urzburg group is funded by the Deutsche Forschungsgemeinschaft (DFG, German Research Foundation) through Project-ID 258499086 - SFB 1170 and through the W\"urzburg-Dresden Cluster of Excellence on Complexity and Topology in Quantum Matter-ct.qmat Project-ID 390858490 - EXC 2147.\\
Y.I. acknowledges financial support by the Science and Engineering Research Board (SERB), Department of Science and Technology (DST), India through the Startup Research Grant No.~SRG/2019/000056, MATRICS Grant No.~MTR/2019/001042, and the Indo-French Centre for the Promotion of Advanced Research (CEFIPRA) Project No. 64T3-1. This research was supported in part by the National Science Foundation under Grant No.~NSF~PHY-1748958, the Abdus Salam International Centre for Theoretical Physics (ICTP) through the Simons Associateship scheme funded by the Simons Foundation, IIT Madras through the Institute of Eminence (IoE) program for establishing the QuCenDiEM group (Project No. SB20210813PHMHRD002720), the International Centre for Theoretical Sciences (ICTS), Bengaluru, India during a visit for participating in the program “Novel phases of quantum matter” (Code: ICTS/topmatter2019/12) and “The 2nd Asia Pacific Workshop on Quantum Magnetism” (Code: ICTS/apfm2018/11). Y.I. acknowledges the use of the computing resources at HPCE, IIT Madras. \\
The numerical simulations were performed on the CHEOPS cluster at RRZK Cologne, the JURECA Booster \cite{jureca} and the JUWELS cluster \cite{JUWELS} at the Forschungszentrum Juelich, the Julia cluster in Würzburg and the GCS Supercomputer SuperMUC-NG at the Leibniz Supercomputing Centre.

\bibliography{bib_mFRG}

\appendix

\onecolumngrid

\section{Half-filling constraint}
\label{appendix:halffilling}

The decomposition of the spin operators $S_{i}$ into auxiliary fermionic partons introduces an artificial enlargement of the Hilbert space, which needs to be handled by a constraint on the occupation number on each lattice site $i$. However, enforcing the constraint exactly, in our case $\sum_{\alpha} f_{i \alpha}^{\dagger} f_{i \alpha}^{\pdagger} = 1$ with $\alpha \in \{\uparrow, \downarrow\}$, is technically difficult, since it would require the inclusion of an additional flowing gauge field in our FRG approach \cite{Roscher}. Therefore, we enforce the constraint only on average $\langle \sum_{\alpha} f_{i \alpha}^{\dagger} f_{i \alpha}^{\pdagger} \rangle = 1$, by imposing particle-hole symmetry on the level of vertices. The successful (numerical) implementation can be checked by computing the product $G_{i \alpha}(\tau) G_{i \alpha}(-\tau)$ for the single-particle Green's function
\begin{align}
	G_{i \alpha}(\tau) = -\langle \hat{T}_{\tau}  f^{\pdagger}_{i\alpha}(\tau) f^{\dagger}_{i \alpha}(0) \rangle \,
\end{align}
in imaginary time, which, written as the convolution of its Fourier transform $G_{i \alpha}(w)$, is given by 
\begin{align}
	G_{i \alpha}(\tau) G_{i \alpha}(-\tau) =\frac{1}{(2 \pi)^2} \int_{-\infty}^{\infty} dw \int_{-\infty}^{\infty} dv \ G_{i \alpha}(w) G_{i \alpha}(w - v) e^{i v \tau}  \,.
	\label{constraint_integral}
\end{align}
If the constraint is fulfilled on average one should then have
\begin{align}
	\lim_{\tau \to 0^{+}} G_{i \alpha}(\tau) G_{i \alpha}(-\tau) = G_{i \alpha}(0^{+}) G_{i \alpha}(0^{-}) = -\frac{1}{4} \,.
	\label{constraint}
\end{align}
For the $T = 0$ implementation of pf-FRG, one cannot directly compute $G_{i \alpha}(0^{\pm})$, since the propagator is an odd function in frequency space, such that an integral over the full frequency domain vanishes. Note that Eq.~\eqref{constraint} should generally hold for any particle-hole symmetric self-energy, especially the one obtained in our FRG flow. Therefore, we have computed the double integral Eq.~\eqref{constraint_integral} with our numerical $\Sigma^{\Lambda}(w)$ as input in order to check the consistency of our implementation. We find that, independent of the scale $\Lambda$, the coupling, the system size and the loop order, the half-filling constraint is indeed fulfilled on average.

\section{Definition of two-particle reducible channels}
\label{appendix:channel_definitions}

In Eq.~\eqref{channeldecomp} we have introduced the decomposition of the two-particle vertex flow in three two-particle reducible channels $\dot{g}_{c}^{\Lambda}$ with $c \in \{pp, dph, cph\}$, which were symbolically defined as 
\begin{align}
    \dot{g}_{c}^{\Lambda} &= \big{[} \Gam \circ \partial_{\Lambda} (G^{\Lambda} \times G^{\Lambda}) \circ \Gam \big{]}_{c} \,.
\end{align}
Starting from Eq.~\eqref{2Pgeneral}, the concrete expressions read 
\begin{align}
    \dot{g}_{pp}^{\Lambda}(1', 2'; 1, 2)  &= -\frac{1}{4\pi} \sum_{3, 4} \G{3}{4}{1}{2}  \ \partial_{\Lambda} (G^{\Lambda}(3) G^{\Lambda}(4)) \ \G{1'}{2'}{3}{4} \\ 
    \dot{g}_{dph}^{\Lambda}(1', 2'; 1, 2) &=  \frac{1}{2\pi} \sum_{3, 4} \G{1'}{4}{1}{3} \ \partial_{\Lambda} (G^{\Lambda}(3) G^{\Lambda}(4)) \ \G{3}{2'}{4}{2}  \\ 
    \dot{g}_{cph}^{\Lambda}(1', 2'; 1, 2) &= -\frac{1}{2\pi} \sum_{3, 4} \G{2'}{4}{1}{3} \ \partial_{\Lambda} (G^{\Lambda}(3) G^{\Lambda}(4)) \ \G{3}{1'}{4}{2}  \,,
\end{align}
where the $pp$ channel needs to be defined with an additional prefactor $\frac{1}{2}$. Note that the crossing symmetry of the two-particle vertex $\G{1'}{2'}{1}{2} = -\G{2'}{1'}{1}{2}$ holds similarly for the $pp$ channel, while the $dph$ and $cph$ channel are exchanged, that is

\begin{align}
    \dot{g}_{pp}^{\Lambda}(1', 2'; 1, 2)  &= -\dot{g}_{pp}^{\Lambda}(2', 1'; 1, 2)  \notag \\ 
    \dot{g}_{dph}^{\Lambda}(1', 2'; 1, 2) &= -\dot{g}_{cph}^{\Lambda}(2', 1'; 1, 2) \notag \\ 
    \dot{g}_{cph}^{\Lambda}(1', 2'; 1, 2) &= -\dot{g}_{dph}^{\Lambda}(2', 1'; 1, 2) \,.
    \label{channel_crossing}
\end{align}

\section{Two-particle reducible channels in bilocal parametrization}
\label{appendix:channel_bilocal}

In Eq.~\eqref{bilocal} we have introduced a bilocal parametrization for the real-space dependence of the two-particle vertex. This representation can be carried over to the two-particle reducible channels from \ref{appendix:channel_definitions} by plugging in the bilocal form and collecting terms with the same spatial structure. This procedure yields
\begin{align}
	\dot{g}_{pp \ i_1 i_2}^{\Lambda}(1', 2'; 1, 2)  &= -\frac{1}{4\pi} \sum_{3, 4}    \Gspatial{i_1}{i_2}{3}{4}{1}{2}  \ \partial_{\Lambda} (G^{\Lambda}(3) G^{\Lambda}(4)) \ \Gspatial{i_1}{i_2}{1'}{2'}{3}{4} \notag \\
	\dot{g}_{dph \ i_1 i_2}^{\Lambda}(1', 2'; 1, 2) &=  \frac{1}{2\pi} \sum_{j, 3, 4} \Gspatial{i_1}{j}{1'}{4}{1}{3}   \ \partial_{\Lambda} (G^{\Lambda}(3) G^{\Lambda}(4)) \ \Gspatial{j}{i_2}{3}{2'}{4}{2}    \notag \\
	                                                &-  \frac{1}{2\pi} \sum_{3, 4}    \Gspatial{i_1}{i_2}{1'}{4}{1}{3} \ \partial_{\Lambda} (G^{\Lambda}(3) G^{\Lambda}(4)) \ \Gspatial{i_2}{i_2}{3}{2'}{2}{4}  \notag \\
	                                                &-  \frac{1}{2\pi} \sum_{3, 4}    \Gspatial{i_1}{i_1}{1'}{4}{3}{1} \ \partial_{\Lambda} (G^{\Lambda}(3) G^{\Lambda}(4)) \ \Gspatial{i_1}{i_2}{3}{2'}{4}{2}  \notag \\
	\dot{g}_{cph \ i_1 i_2}^{\Lambda}(1', 2'; 1, 2) &= -\frac{1}{2\pi} \sum_{3, 4}    \Gspatial{i_1}{i_2}{3}{2'}{1}{4} \ \partial_{\Lambda} (G^{\Lambda}(3) G^{\Lambda}(4)) \ \Gspatial{i_1}{i_2}{1'}{4}{3}{2}  \,.          
	\label{bubbledef}
\end{align}
Here, the multi-indices on the right hand side only contain a spin and frequency index, with spatial indices written out explicitly. Vertices $\Gam$ are to be understood as a short-hand notation for the bilocal vertex component with parallel legs, i.e. $\Gamma^{\Lambda =}$. For the local vertices $\Gam_{i_1 i_1}$ and $\Gam_{i_2 i_2}$ in the second and third term of the $dph$ channel, crossing symmetry was applied to map $\Gamma^{\Lambda \times}$ to $\Gamma^{\Lambda =}$. This is irrelevant as long as full vertices $\Gam$ are used in this expression. For the ml-FRG flow Eq.~\eqref{mFRG_scheme}, however, one also needs to insert only partial contributions to the full vertex. In this case, the channel mapping \ref{channel_crossing} needs to be accounted for explicitly.

\section{Symmetries of two-particle reducible channels}
\label{appendix:symmetries}

In previous work \cite{BuessenOffDiag} a full symmetry analysis for the two-particle vertex in the presence of non-diagonal spin interactions has been performed. Although we focus our effort on Heisenberg spin systems here, we may nevertheless use the derived symmetries in the special case of diagonal interactions. To this end we use the $SU(2)$ symmetric parametrization of the bilocal vertex into a spin (\textit{s}) and density (\textit{d}) component 
\begin{align}
	\Gspatial{i_1}{i_2}{1'}{2'}{1}{2} = \bigg{[} \Gamma^{\Lambda s}_{i_1 i_2}(s, t, u) \sum_{\mu} \sigma^{\mu}_{\alpha_{1'} \alpha_1} \sigma^{\mu}_{\alpha_{2'} \alpha_2} + \Gamma^{\Lambda d}_{i_1 i_2}(s, t, u) \delta_{\alpha_{1'} \alpha_1} \delta_{\alpha_{2'} \alpha_2}\bigg{]} \delta(w_{1'} + w_{2'} - w_1 - w_2) \,,
\end{align}
for which the symmetries read
\begin{align}
	\Gamma^{\Lambda s/d}_{i_1 i_2}(s, t, u) &= \Gamma^{\Lambda s/d}_{i_2 i_1}(-s, t, u) \\
	\Gamma^{\Lambda s/d}_{i_1 i_2}(s, t, u) &= \Gamma^{\Lambda s/d}_{i_1 i_2}(s, -t, u) \\
	\Gamma^{\Lambda s/d}_{i_1 i_2}(s, t, u) &= \Gamma^{\Lambda s/d}_{i_2 i_1}(s, t, -u) \\
	\Gamma^{\Lambda s/d}_{i_1 i_2}(s, t, u) &= \Gamma^{\Lambda s/d}_{i_1 i_2}(u, t, s)  \,,
\end{align}
where $\zeta = +1$ for the spin part and $\zeta = -1$ for the density part. Combinations of one or more symmetries can directly be related to symmetries of the channels by recalling, that the fermionic frequencies of each channel are directly related to linear combinations of the three bosonic transfer frequencies.
This yields
\begin{align}
	g^{\Lambda s/d}_{pp \ i_1 i_2}(s, v_s, v'_s) &= g^{\Lambda s/d}_{pp \ i_2 i_1}(-s, v_s, v'_s) \\
	g^{\Lambda s/d}_{pp \ i_1 i_2}(s, v_s, v'_s) &= \zeta g^{\Lambda s/d}_{cph \ i_2 i_1}(s, -v_s, v'_s) \\
	g^{\Lambda s/d}_{pp \ i_1 i_2}(s, v_s, v'_s) &= \zeta g^{\Lambda s/d}_{cph \ i_1 i_2}(s, v_s, -v'_s) \\
	g^{\Lambda s/d}_{pp \ i_1 i_2}(s, v_s, v'_s) &= g^{\Lambda s/d}_{pp \ i_2 i_1}(s, v'_s, v_s) \,,
\end{align}
for the \textit{pp}-channel,
\begin{align}
	g^{\Lambda s/d}_{dph \ i_1 i_2}(t, v_t, v'_t) &= g^{\Lambda s/d}_{dph \ i_1 i_2}(-t, v_t, v'_t) \\
	g^{\Lambda s/d}_{dph \ i_1 i_2}(t, v_t, v'_t) &= \zeta g^{\Lambda s/d}_{dph \ i_1 i_2}(t, -v_t, v'_t) \\
	g^{\Lambda s/d}_{dph \ i_1 i_2}(t, v_t, v'_t) &= \zeta g^{\Lambda s/d}_{dph \ i_1 i_2}(t, v_t, -v'_t) \\
	g^{\Lambda s/d}_{dph \ i_1 i_2}(t, v_t, v'_t) &= g^{\Lambda s/d}_{dph \ i_2 i_1}(t, v'_t, v_t) \,,
\end{align}
for the \textit{dph}-channel, and finally
\begin{align}
	g^{\Lambda s/d}_{cph \ i_1 i_2}(u, v_u, v'_u) &= g^{\Lambda s/d}_{cph \ i_2 i_1}(-u, v_u, v'_u) \\
	g^{\Lambda s/d}_{cph \ i_1 i_2}(u, v_u, v'_u) &= \zeta g^{\Lambda s/d}_{pp \ i_2 i_1}(u, -v_u, v'_u) \\
	g^{\Lambda s/d}_{cph \ i_1 i_2}(u, v_u, v'_u) &= \zeta g^{\Lambda s/d}_{pp \ i_1 i_2}(u, v_u, -v'_u) \\
	g^{\Lambda s/d}_{cph \ i_1 i_2}(u, v_u, v'_u) &= g^{\Lambda s/d}_{cph \ i_1 i_2}(u, v'_u, v_u) \,,
\end{align}
for the \textit{cph}-channel. Given these symmetries, one can further conclude how they affect the respective kernel functions by successively eliminating certain kernels considering their asymptotic limit. Performing the full symmetry analysis we were able to drastically reduce the numerical effort in computing the two-particle vertex flow. Most notably, all kernels need to be saved only for positive Matsubara frequencies. Finally the $v \leftrightarrow v'$ symmetry allows to restrict $Q^{s/d}_{3 \ c}$ to a mesh with $v \geq v'$.

\section{Scanning routine for frequency mesh adaptation}
\label{appendix:scanning}

Continuing the ml-FRG flow to small values of the flow parameter requires that all relevant features of the vertices are well resolved in intermediate stages of the flow. Carefully analysing the vertices we found that most structure is usually located around the zero frequency regime, where sharp peaks right at or close to $(w_c, v_c, v'_c) = (0, 0, 0)$ appear, and for lattice sites close to the reference site $i_0$. Our routine to scan the vertices after each step of the ODE solver and to determine from that a suitable linear spacing $h$ for the frequency meshes, uses the relative deviation $\Delta = \frac{|g_2 - g_1|}{\mathrm{max}(|g_2|, |g_1|)}$ as a control parameter. Here $\{ g_i \}$ are the respective vertex values along a given frequency axis $\{ w_i \}$ with $w_1 = 0$. More precisely, the mesh spacing $h$ is increased or decreased such that $p_1 \leq \Delta \leq p_2$, where $p_1$ and $p_2$ are external parameters. We choose $p_1 = 0.05$ and $p_2 = 0.1$. As an additional sanity check, the spacing $h$ must fulfill $p_3 \Lambda \leq h \leq p_4 \Lambda$, to avoid overambitious shrinking or expanding of the linear part of the mesh. We choose $p_3 = 0.05$ and $p_4 = 2.0$. The scanning is carried out for the bosonic and fermionic axis right at the reference site $i_0$ for all channels and with the respective other frequency arguments set to zero. Note, that this scanning is only carried out when $\textrm{max}(\{|g_i|\}) > 10^{-3}$ to prevent adapting the meshes according to noisy (i.e. not well captured with respect to the chosen error tolerances) data.

\end{document}